%% file: NEOWISEpaper.tex
\documentclass[preprint]{aastex631}

\newcommand\aastex{AAS\TeX}

\accepted{for publication in The Astronomical Journal}

\shorttitle{\aastex\ WISE/NEOWISE for Phaethon, 2005~UD and 1999~YC}
\shortauthors{Kasuga \& Masiero (2022)}

\usepackage{mediabb}
\usepackage{amsmath}
\usepackage{amssymb}
\usepackage{rotating, graphicx}
\usepackage{longtable}
\usepackage[stable]{footmisc}
\usepackage{mdwlist}

\begin{document}

\title{WISE/NEOWISE Multi-Epoch Imaging of the Potentially Geminid-related Asteroids: \\
(3200) Phaethon, 2005~UD and 1999~YC}

\correspondingauthor{Toshihiro Kasuga}
\email{toshi.kasuga@nao.ac.jp}

\author[0000-0001-5903-7391]{Toshihiro Kasuga}
\affil{National Astronomical Observatory of Japan, 2-21-1 Osawa, Mitaka, Tokyo 181-8588, Japan}

\author[0000-0003-2638-720X]{Joseph R. Masiero}
\affiliation{Caltech/IPAC, 1200 E California Blvd, MC 100-22, Pasadena, CA 91125, USA}



\begin{abstract}

We present space-based thermal infrared observations of  
the presumably Geminid-associated asteroids: 
(3200)~Phaethon, 2005~UD and 1999~YC 
using ${\it WISE}$/NEOWISE.  
The images were taken at the four wavelength bands 
3.4\micron~($W1$), 4.6\micron~($W2$), 12\micron~($W3$), 
and 22\micron~($W4$).  
We find no evidence of lasting mass-loss 
in the asteroids over the decadal multi-epoch datasets.  
We set an upper limit to the mass-loss rate in dust of
$Q_{\rm dust}$~$\lesssim$~2\,kg\,s$^{-1}$ for Phaethon 
and $\lesssim$~0.1\,kg\,s$^{-1}$ for both 2005~UD and 1999~YC, respectively,  
with little dependency over the observed heliocentric distances of $R_{\rm h}$=1.0--2.3\,au.    
For Phaethon, even if the maximum mass-loss was sustained over the 1000(s)\,yr 
dynamical age of the Geminid stream, it is more than two orders of magnitude 
too small to supply the reported stream mass (10$^{13-14}$\,kg).   
The Phaethon-associated dust trail (Geminid stream) is not detected at $R_{\rm h}$=2.3\,au, 
corresponding an upper limit on the optical depth of $\tau$~$<$~7$\times$10$^{-9}$. 
Additionally, no co-moving asteroids with radii $r_{\rm e} <$~650\,m were found.   
The DESTINY$^{+}$ dust analyzer would be capable of detecting 
several of the 10\micron-sized interplanetary dust particles 
when at far distances ($\gtrsim$50,000\,km) from Phaethon.     
From 2005~UD, if the mass-loss rate lasted 
over the 10,000\,yr dynamical age of the Daytime Sextantid meteoroid stream, 
the mass of the stream would be $\sim$10$^{10}$\,kg. 
The 1999~YC images showed neither the related dust trail (the optical depth $\tau <$ 2$\times$10$^{-8}$) 
nor co-moving objects with radii $r_{\rm e} <$~170\,m at $R_{\rm h}$=1.6\,au.  
Estimated physical parameters from these limits do not explain the 
production mechanism of the Geminid meteoroid stream.
Lastly, to explore the origin of the Geminids, we discuss 
the implications for our data in relation to 
the possibly Sodium (Na)-driven perihelion activity of Phaethon.    

\end{abstract}





\section{Introduction} 
\label{intro}

The near-Earth asteroid (NEA) (3200) Phaethon (hereafter just Phaethon) 
appears to be dynamically associated with 
the Geminid meteoroid stream
$\footnote{GEM/4 from IAU Meteor Data Center, 
Nomenclature \citep{JopekJenniskens11}}$ \citep{Whipple1983,Gu89,WW93}. 
The semimajor axis $a$=1.271\,au, eccentricity $e$= 0.890, and inclination $i$=$22^{\circ}.3$ 
(NASA JPL HORIZONS$\footnote{\url{https://ssd.jpl.nasa.gov/horizons/}}$) 
corresponds to a Tisserand parameter $T_J$ = 4.5, which is 
distinctly above those of cometary orbits ($T_J \leqslant$ 3.08) 
and is classified as having a typical asteroidal orbit \citep{JewittHsieh2022}.  
The Geminid meteoroid stream consists of near millimeter-scale 
or larger solid particles \citep[by radar,][]{Bla17}, up to 10\,s centimeters, 
as measured by lunar impact flushes 
\citep[][]{Yanagisawa08,Yanagisawa2021,Szalay18,Madiedo2019msme}.  
The remarkable orbital feature is the small perihelion, $q$ = 0.14\,au, 
where Phaethon is repeatedly exposed to intense 
thermal processing at the peak sub-solar temperature $\gtrsim$ 1000\,K.     

Spectroscopic measurements of sodium (Na) content of the Geminid meteors 
have been utilized for studying the thermal processes on Phaethon.  
The brightness of obtained spectra are at optical absolute magnitudes $+$4 or brighter, 
corresponding to meteoroid sizes $\gtrsim$ 1\,mm \citep[e.g.][]{Lindblad1987}. 
Sodium is relatively volatile metal element in meteoroids in 
most meteor showers, and commonly detected as 
neutral Na--D emission lines (Na\,{\sc i}--doublet) at the wavelength $\lambda$$\sim$5890\,\AA~\citep{CBE98}.  
But, a notable feature of the Geminid meteors is that the extreme variety in their Na content, 
from depletion of the Na abundance ($\sim$7\% of the solar) \citep{KasugaGeminid2005} 
to near the solar-like values \citep[e.g.][]{Harvey73,B01,TLBF03}.  
Similar trends are reported from the measured intensity ratios of neutral metallic atom emission lines, 
as well as undetected Na\,{\sc i} \citep{B01,Borovicka2005,Borovicka2019msme}.  
\cite{Kasuga2006AA} compiled Na in the Geminids in the last decade 
to investigate perihelion-dependent thermal effects on meteoroid streams.   
The effect is predicted to alter the metal abundances from their intrinsic values in their parents, 
especially for temperature-sensitive elements such as Na in alkaline silicates.  
However, as a result, 
the thermal desorption of Na is unlikely to occur 
for the Geminid stream even at $q = 0.14$\,au. 
This is because the corresponding meteoroid temperature 
(characterized as large, compact, blackbody-like particle) 
does not reach the sublimation temperature of alkaline silicates \citep{Kasuga2006AA}\citep[also see,][]{Springmann19}. 
The mm-scale or larger Geminid meteoroids take timescales of~$\gtrsim$~10$^{4\sim5}$\,yr  
to lose Na \citep{CapekBorovicka2009}, 
being 1--2 orders of longer than the stream age $\tau_s \sim$ 1000(s)\,yr 
\citep{WW93,Ryabova99,JakubNeslu15} \citep[Reviewed in][]{Vaubaillon19}.  
Therefore the Na loss observed in the Geminid meteors must have originated                
from the thermal process of the parent, Phaethon \citep[see a review, ][]{KasugaJewitt2019}.

The parent body Phaethon has a diameter $D_{\rm e}$$\approx$\,5--6\,km \citep{Tedesco04,Hanus2016AA,Taylor2019PSS,Dunham2019LPI,Masiero2019AJ,Devog2020PSJ}\footnote{ 
The effective spherical diameter $D_{\rm e}$$\sim$\,5.5\,km (a mean equatorial diameter of 6.3\,km) is estimated by the radar \citep{Taylor2019PSS} 
and $D_{\rm e}$=5.2$\pm$0.1\,km is by the occultation \citep{Dunham2019LPI,Devog2020PSJ}, respectively.  
These are the least model dependent.  
} 
and an optically blue (B-type) reflection spectrum \citep[e.g.][]{BusBinzel2002Icar}.  
While most images appear as a point source without any coma \citep{CMSS96,HsiehJewitt2005,Wiegert2008Icar}, 
recurrent dust-releasing activity has been optically observed only at the perihelion passage 
\citep[][]{JewittLi10,Jewitt12,Jewitt13ApJ,Li_Jewitt2013AJ,HuiLi17}.
This is interpreted as a thermal-induced activity.   
The proposed mechanisms are thermal breakdown \citep{JewittLi10,Jewitt12}, 
mass-shedding by thermal deformation \citep{Nakano2020ApJ}, 
gas-dragged (or -gradient) force by sublimation of sodium \citep{Masiero2021PSJ},  
and electrostatic repulsive forces \citep{Jewitt12,Jewitt_active15} 
which might be caused by accumulated sodium ions \citep{Kimura2022arXiv}.  
Constraints on mechanisms are provided both by the size of ejected particles and the mass-loss rates.    
The released particles are of micron size (measured from the short tail), 
too small compared to the millimeter$\sim$centimeter-sized Geminids.  
Such small particles would not be retained in the stream due to 
radiation pressure \citep[Reviewed in][]{Jewitt_active15,JewittHsieh2022}.   
Larger particles (sizes $>$ 10$\micron$) must be in the Geminid stream (trail) 
as detected by the $DIRBE/COBE$ thermal infrared survey from space \citep{Arendt14}.   
Furthermore, even if the measured perihelion mass-loss rates $\sim$3\,kg\,s$^{-1}$ \citep{JewittLi10}
sustained in steady-state over the entire orbit, it is orders of magnitude too small 
to supply the Geminid stream mass $M_s$ $\approx$ $10^{13}$\,kg  -- $10^{14}$\,kg \citep{HM89,Bla17} 
\citep[cf. $\lesssim 10^{15}$\,kg if cometary origin, ][]{Ryabova17} 
within its $\tau_s \sim$ 1000(s) years of dynamical age.  
The observed particle size and mass-loss rate from Phaethon are not consistent with the Geminid stream, 
leading to the open question that how the stream was produced.   

A non-negligible possibility is that the Geminids are the products of 
a catastrophic event (e.g. collision, disruption).  
Phaethon has a hypothetical breakup event which might occurred $>$~10$^4$\,yr ago.   
This produced fragmental kilometer-sized NEAs: 
(155140)~2005~UD and (225416)~1999~YC (hereafter 2005~UD and 1999~YC, respectively), 
named as ``Phaethon-Geminid Complex (PGC)" \citep[][]{OSK06,OAI08}.  
The optical color of 2005~UD is blue (B-type), consistent with Phaethon \citep{JewittHsieh2006}, 
but the dissimilarity in the concave-shaped near-infrared spectrum 
makes a direct association debatable \citep{Kareta2021PSJ,MacLennan2022arXiv}.  
The color of 1999~YC in the visual band distinctly exhibits a neutral (C-type) slope \citep{KasugaJewitt2008AJ}.  
The PGC-asteroids show some physical similarity, 
but their dynamical association is unclear \citep[see a review,][]{KasugaJewitt2019}.  
Relevantly, Phaeton is proposed to have its precursor origin in the main-belt. 
The dynamical lifetime of 26\,Myr can be linked back to 
(2)~Pallas \citep{deLeon10} with $\sim$ 2\% chance \citep{NT18} 
or to the inner main-belt asteroids (e.g. (329)~Svea or (142)~Polana) 
with $\sim$ 18\% chance \citep{MacLennan2021Icar}.  
The links are of very low probabilities and still unsure, but imply 
that Phaeton itself could be a disrupted fragment from one of these sources.  
To summarize, the Geminid stream and Phaethon, both 
are likely to be formed stochastically rather than in steady-state, 
however the nature of the events are not clarified yet.

In this paper we present a space-based thermal infrared study of 
the Geminid-related NEAs: Phaethon, 2005~UD and 1999~YC 
using the NASA spacecraft {\it Wide-field Infrared Survey Explorer} \citep[WISE;][]{Wright2010AJ} 
and the successive project {\it Near-Earth Object WISE} \citep[NEOWISE;][]{Mainzer2011ApJ, Mainzer2014ApJ}.   
Long wavelength observations are aimed at larger 
particles potentially containing much more mass. 
The multiple observing epochs contain data from four infrared wavelength bands: 
3.4$\micron$ ($W1$), 4.6$\micron$ ($W2$), 12$\micron$ ($W3$), and 22$\micron$ ($W4$),
providing opportunities to constrain the dust environments around 
Phaethon, 2005~UD and 1999~YC over the last decade.  
The each band is specified for investigating different physical characters of 
the PGC-asteroids, such as the dust production rate ($W1$), 
the gas production rate ($W2$) and dust trails 
and potentially co-moving objects ($W3$ and $W4$). 
Another motivation is to find an activity mechanism far from perihelion,  
which is different from thermal-related production.   
This study provides a groundwork for JAXA's $DESTINY^{+}$ mission which is  
planning to flyby at the distance of 500\,km from Phaethon in 2028, 
with a potential extended mission to flyby 2005~UD \citep{Ozaki2022arXiv}.

\section{{\it WISE}/NEOWISE Data}
\label{wise}

The near- and mid-infrared data were taken 
by ${\it WISE}$ \citep[][]{Wright2010AJ} 
and NEOWISE \citep[][]{Mainzer2011ApJ, Mainzer2014ApJ}.  
${\it WISE}$ is equipped with a 40\,cm diameter telescope, which was 
launched on 2009 December 14 into a Sun-synchronous polar orbit.
The full sky survey was conducted using four infrared wavelength 
bands centered at 3.4$\micron$ ($W1$), 4.6$\micron$ ($W2$), 
12$\micron$ ($W3$), and 22\,$\micron$ ($W4$).
It was continued until 2010 September 29 when the cryogen was exhausted. 
Then ${\it WISE}$ further continued surveying
in a post-cryogenic two-band survey mode 
until 2011 February 1 when it was put into hibernation.
On 2013 December 13, NEOWISE resumed the two-band survey 
($W1$ and $W2$), which is currently ongoing as a reactivation mission (NEOWISE-R).  

The spacecraft employs two-types of focal plane array detectors, 
the HgCdTe (Teledyne Imaging Sensors) in the shorter wavelength bands ($W1$ and $W2$) 
and the Si:As BIB (DRS Sensors \& Targeting Systems) in 
the longer wavelength bands ($W3$ and $W4$) \citep{Wright2010AJ}.  
Both detectors have 1024~$\times$~1024 pixels, 
each $2\arcsec.76$ pix$^{-1}$ in the $W1$, $W2$ and $W3$ bands 
and $5\arcsec.5$ pix$^{-1}$ (2$\times$2 binned) in the $W4$ band.  
The corresponding field of view (FOV) is 47$\arcmin$$\times$47$\arcmin$.
With the dichroic beam splitters, consecutive frames were simultaneously 
imaged through the same FOV in each band every 11\,second.  
The exposure time was 7.7\,seconds in the W1 and W2 bands 
and 8.8\,seconds in the $W3$ and $W4$ bands, respectively \citep{Wright2010AJ}.  
The data of {\it WISE} ($W1$--$W4$) and NEOWISE ($W1$ and $W2$) are available 
at the NASA/IPAC Infrared Science Archive (IRSA\footnote{\url{https://irsa.ipac.caltech.edu/frontpage/}}).   
The instrumental, photometric, and astrometric calibrations are detailed in \cite{Cutri2012wise.rept, Cutri2015nwis.rept}.    

We used the IRSA search tools (GATOR\footnote{\url{https://irsa.ipac.caltech.edu/cgi-bin/Gator/nph-scan?submit=Select&projshort=WISE}} 
and WISE Image Service\footnote{\url{https://irsa.ipac.caltech.edu/applications/wise/}}) 
to retrieve all detections of the three PGC asteroids:  Phaethon, 2005~UD and 1999~YC.  
In the GATOR, the queried sources are 
WISE All-Sky Single Exposure (L1b) \citep{https://doi.org/10.26131/irsa139}, 
WISE 3-Band Cryo Single Exposure (L1b) \citep{https://doi.org/10.26131/irsa127}, 
WISE Post-Cryo Single Exposure (L1b) \citep{https://doi.org/10.26131/irsa124}, 
and 
NEOWISE-R Single Exposure (L1b) \citep{https://doi.org/10.26131/irsa144}.  
The detections were double-checked with the WISE Image Service, 
and compared with the list from 
the Minor Planet Center (MPC) database\footnote{\url{https://minorplanetcenter.net/db_search}}.   
Here, we applied the screening process to the data described in \cite{Masiero2011ApJ,Hung2022}.  
We based our selection criteria on 
high Signal-to-Noise Ratio (SNR) ($>$~5) in at least one band, 
moon angular separation $>$~30\,degrees, 
and angular distance from the nominal boundaries of 
the South Atlantic Anomaly (SAA) $\gtrsim$~10\,degrees.  
We selected observations having flags of \verb|cc_flags| = 0 
which secures no contamination produced by known artifacts, 
e.g. latent images or diffraction spikes \citep{Mainzer2011ApJ}.
We used the profile-fit photometric quality flag of \verb|ph_qual| = A, B, C or U 
which corresponds to SNR~$\ge$~10, 3~$<$~SNR~$<$~10, 2~$<$~SNR~$<$~3 
and SNR~$\leq$ 2 (= 2$\sigma$ upper limit on magnitude, see also \S~\ref{magflux}), respectively.    
The highest frame quality score of \verb|qual_frame|=10 is used.   

The ${\it WISE}$/NEOWISE data reduction pipeline conducts 
initial nonlinearity and saturation correction for data brighter than 
the threshold of $W1$ = 8.1\,mag, $W2$ = 6.7\,mag, $W3$ = 3.8\,mag, and $W4$ = $-$0.4\,mag, however further analysis shows that additional corrections are needed 
\citep{Cutri2012wise.rept, Cutri2015nwis.rept}\footnote{\url{https://wise2.ipac.caltech.edu/docs/release/allsky/expsup/sec6_3d.html}}.  
For Phaethon taken on MJD~58104.2148 (UT 2017-Dec-17),  
the measured magnitudes listed in the catalog, 
$W1$=7.745\,$\pm$\,0.014\,mag and $W2$=2.894\,$\pm$\,0.003\,mag, 
are in the saturated regime.  
We corrected their saturation biases using the relations 
derived for those of specific catalog magnitudes
\footnote{\url{https://wise2.ipac.caltech.edu/docs/release/neowise/expsup/sec2_1civa.html}} \citep[cf.][]{Masiero2019AJ},   
\begin{eqnarray}
\begin{split}
W1({\rm out}) & = &  W1({\rm in}) + (0.037\pm0.039)\,{\rm mag},  \\
W2({\rm out}) & = &  W2({\rm in}) + (1.330\pm0.221)\,{\rm mag}.
\end{split}
\label{saturated}
\end{eqnarray}
We substitute $W1({\rm in})$=7.745$\pm$0.014\,mag and $W2({\rm in})$=2.894$\pm$0.003\,mag into Equation~(\ref{saturated}), 
and we obtain the corrected magnitude $W1({\rm out})$=7.782$\pm$0.041\,mag and $W2({\rm out})$=4.224$\pm$0.221\,mag, respectively.

We visually checked the image data to remove star contamination 
that might appear in the shorter bands ($W1$ and $W2$) even 
if they do not appear in the longer wavelength bands ($W3$ and $W4$).
This is because the stellar thermal signatures ($\gtrsim$~2000\,K) 
have their blackbody peaks in the shorter wavelengths 
and have flat decreasing spectra, representing $W1$~SNR~$>$~$W2$~SNR. 
Likewise, as for the two-band detections of NEOWISE, 
we removed the contamination by requiring
$W1-W2$ $>$~1~mag.  
The principle is more detailed in \cite{Masiero2018AJ, Masiero2020PSJ}. 
We actually find this issue in 1999~YC taken by ${\it WISE}$ on 
MJD 55213.7089 (UT 2010-Jan-17), requiring us to drop its shorter bands.  
Finally we again visually conducted spot check on all of the selected data 
to ensure no corruptions in the images. 
The image corrupted by artifacts and noises are avoided.     
We found five observation epochs for Phaethon, three for 2005~UD and two for 1999~YC.  
The observation log is shown in Table~\ref{obslog}. 
The orbital information is summarized in Table~\ref{PGCorbits}. 

Next, we inspected the full-width half maximum (FWHM) of asteroid images 
to find if the surface brightness profiles can be used to set their activity limits \citep{Jewitt19Thermal}.  
We here show an example of Phaethon's point-spread function (PSF) 
taken in the $W3$ band on MJD~55203.2564 (UT 2010-Jan-07). 
The FWHM of Phaethon is $\theta_{\rm F}$=$7\arcsec.8$, beyond 
the diffraction limit of the {\it WISE} instrument $\theta_{\rm d}$=$6\arcsec.4$ 
(=1.03\,$\lambda$/$D$ in radians, where $\lambda$=12$\micron$ is 
the wavelength band center and $D$=40\,cm is the telescope diameter).  
For comparison, the FWHM of nearby field stars is $\sim$$6\arcsec.6$ (medium value of seven field stars).  
The angular diameter of Phaethon (a diameter $D_{\rm e}$ $\sim$ 5\,km at 
$\Delta \sim$ 2.08\,au) subtends $\theta_{\rm Ph}$=$0\arcsec.0031$, 
which is negligibly small.  
The derived composite width of Gaussian is $6\arcsec.4$  
(= $(\theta_{\rm d}^2$ + $\theta_{\rm Ph}^2$)$^{0.5}$), 
which is not sufficient to resolve the Phaethon image.  
The same circumstance is found for the other data of interest.  
Therefore, the surface brightness profiles are not expected to give useful limits 
to their activity levels.     

In this study, we use the effective spherical diameters ($D_{\rm e}$) 
and geometric visible albedos ($p_{\rm v}$) determined 
from ${\it WISE}$/NEOWISE (Table~\ref{Size_pv}).        
Many observing-epochs cover multiple 
viewing geometries, producing 
the refined data sets in the uniformity and calibration.  
Phaethon, for example, five epochs of data 
estimated $D_{\rm e}$=4.6$^{+0.2}_{-0.3}$\,km by applying   
Thermophysical model (TPM) fits \citep{Masiero2019AJ}.  
The accuracy is validated by comparing the estimated sizes 
of 23 main-belt asteroids (surface temperature variation is stable) from NEOWISE 
with those from occultation/spacecraft as the ground-truth sizes \citep{Masiero2019AJ}.  
The size is consistent within 2$\sigma$ of the other TPM result 
$D_{\rm e}$=5.1$\pm$0.2\,km \citep{Hanus2016AA,Hanus2018AA}, 
while not consistent with $D_{\rm e}$$\sim$5.5\,km showing 
a muffin-top shape (equatorial bulge of 6.3\,km) as observed 
by Arecibo radar during the apparition in 2017-Dec \citep{Taylor2019PSS}. 
The shape model or observed orbital positions (pre- or post-$q$) may 
require some offsets in size estimations and thermophysical behavior \citep{MacLennan2021PSJ,MacLennan2022arXiv}. 
Here, we describe the justification for applying 
the ${\it WISE}$/NEOWISE-derived size with its spherical assumption.   
The thermal survey and radar data have a huge difference in the resolution, 
for example 140\,km\,pix$^{-1}$ (NEOWISE) vs. 0.075\,km\,pix$^{-1}$ (Arecibo) 
at the Phaethon's closest distance ($\Delta$=0.069\,au).  
The infrared data do not resolve a 5-6\,km asteroidal shape, indicating 
that the assumption of spherical body is reasonable for the usage of ${\it WISE}$/NEOWISE alone.  
The post-$q$ data tend to estimate larger sizes than those of pre-$q$ data
due to significant bias suffered from extreme heating at its perihelion passage \citep{MacLennan2022arXiv}.  
In such a case, the pre-$q$ data use a stable variation of surface temperature 
presumed to be plausible for obtaining 
the statistically best-fitted size $D_{\rm e}$=4.8$\pm$0.2\,km \citep[Table~4 in][]{MacLennan2022arXiv}.  
This is consistent within 1$\sigma$ of our ${\it WISE}$/NEOWISE-derived size.    
On the other hand, 
the km-scale asteroids with a few epochs of data find that 
the TPM-fitted size is consistent within $\lesssim$\,0.1\,km  
of the other model \citep{Masiero2019AJ}.     
Thus we apply the unified uncertainty in Phaethon size, 
i.e. $D_{\rm e}$=4.6$\pm$0.3\,km for simplicity of use.

\section{Analysis}

\subsection{Photometry, Calibration and Conversion from Magnitude to Flux Density}
\label{magflux}

The photometry and calibration methodology are given 
by the IRSA website\footnote{\url{https://wise2.ipac.caltech.edu/docs/release/allsky/expsup/sec4_4h.html}} \citep{Cutri2012wise.rept, Cutri2015nwis.rept}. 
Magnitudes are measured based on the median FWHM of the PSF of the asteroid at the each band.
The aperture radius is set as 1.25~$\times$~FWHM, where the FWHM adopts 
6$\arcsec$, 6$\arcsec$, 6$\arcsec$, and 12$\arcsec$ in bands W1 through W4, respectively\footnote{\url{https://wise2.ipac.caltech.edu/docs/release/allsky/expsup/sec4_4c.html}}.
When the Profile-fitting Photometry (WPRO) or Aperture Photometry System (WAPP) measure 
the flux with the SNR $\leq$ 2, magnitude uncertainty is expressed as ``NULL" 
and the calibrated magnitude is interpreted as the 2$\sigma$ upper limit\footnote{\url{https://wise2.ipac.caltech.edu/docs/release/allsky/expsup/sec4_4c.html\#ul2}}.
Those results are listed without uncertainties in Table~\ref{obslog}.   

The measured magnitudes are converted into the flux density (Jy) using the published zero points 
and the color corrections \citep{Wright2010AJ}.  
The derived flux density can be converted to SI units (W~m$^{-2}$~$\micron$$^{-1}$)  
using the relative system response curves (RSRs) and the zero magnitude attributes \citep[Table~1, ][]{Jarrett2011ApJ}.  
The measured instrumental source brightness (in digital numbers) is calibrated by referring 
an instrumental zero point magnitude\footnote{\url{https://wise2.ipac.caltech.edu/docs/release/allsky/expsup/sec4_4h.html\#CalibratedM}} 
\citep{Cutri2012wise.rept, Cutri2015nwis.rept}.  
Hereafter we apply the procedure as appropriate.

\subsection{Reflected Sunlight Removal}
\label{Ref}

The asteroids from ${\it WISE}$/NEOWISE comprise 
a blend of reflected sunlight and thermal emission.  
It is significant in the shorter bands ($W1$ and $W2$) of km-sized NEAs \citep{Mainzer2011ApJ,Nugent2015ApJ}.
Here we remove the reflected sunlight to extract thermal emission alone.  

The flux density of reflected sunlight from an asteroid, 
$F^{\rm Ref}_\nu$~(Jy), 
is calculated correcting the Sun-observer-object distance for each frame, using the equation  
\begin{equation}
F^{\rm Ref}_\nu  =  \frac{\pi\,B_\nu (T_\odot)\,R_\odot^2}{R_{h}^2}\,A\,\pi\,\left(\frac{D_{\rm e}}{2}\right)^2\,\frac{\Phi_{\rm vis}(\alpha)}{\Delta^2}, 
\label{Fvis}
\end{equation}
where $B_\nu$ is the Planck function (Jy~sr$^{-1}$) at the Solar temperature $T_\odot$ = 5778\,K, 
$R_\odot$ = 6.957 $\times$10$^{10}$~cm is the Solar radius, 
$R_{\rm h}$ is heliocentric distance (cm).  
The $A \approx A_{\rm v}$ = $q\,p_{\rm v}$ is the bolometric Bond albedo (scattered fraction of the incident solar flux),
where $q$ is the phase integral (measure of the angular dependence of the scattered emission), 
and 
$p_{\rm v}$ is the visible geometric albedo.  
Based on the shape of phase curve in the $H-G$ model of \cite{Bowell1989aste.conf}, 
we adopt $q$=0.3926 from $q$=0.29+0.684$G$ where $G$=0.15 is the phase slope parameter.   
$D_{\rm e}$ is the asteroid diameter (cm), 
$\Phi_{\rm vis}(\alpha)$ is the phase function \cite[Equation~(A4) of][]{Bowell1989aste.conf} in which 
$\alpha$ is the phase angle (in degrees),  
and 
$\Delta$ is the ${\it WISE}$-centric distance (cm).
Using the measured values of $D_{\rm e}$ and $p_{\rm v}$ from {\it WISE}/NEOWISE (Table~\ref{Size_pv}), 
we obtained $F^{\rm Ref}_\nu$ to remove from the measured flux.   
The extracted thermal flux density is summarized in Table~\ref{thermalflux}.

\subsection{Temperature}

In order to estimate an effective temperature, $T_{\rm eff}$~(K),  
we use the thermal flux densities at the $W3$ (12\,$\micron$) and $W4$ (22\,$\micron$) from {\it WISE}.   
The shorter wavelength bands ($W1$ and $W2$) 
are avoided to prevent any contamination from remaining blended sources.
The ratio of flux density, $W3$/$W4$, corresponds to 
the ratio calculated from a blackbody having $T_{\rm eff}$ \citep{Jewitt19Thermal}.
As a reference, an equilibrium blackbody temperature ($T_{\rm bb}$) of a sphere, 
isothermal object at the same heliocentric distance ($R_{\rm h}$) is given. 
The results are shown in Table~\ref{Teff}.  

The derived effective temperature is close to blackbody temperature, 
$T_{\rm eff}$~$\approx$~0.98$\times$$T_{\rm bb}$.   
This suggests that Phaethon and 1999~YC are approximately 
considered as sphere, blackbody-like objects in the {\it WISE}/NEOWISE data.  
This interpretation is different from the previous high-resolved thermal image of Phaethon. 
During its 2017 flyby, the $VLT$ obtained the excess effective temperature ($T_{\rm eff}$ $\approx$ 1.14$\times$$T_{\rm bb}$), 
implying that Phaethon ($D_{\rm e} \sim$5\,km) has a non-spherical, non-blackbody features in detail \citep[][]{Jewitt19Thermal}.  
The difference in $T_{\rm eff}$ could be caused by the difference in the observed phase angle.  
While {\it WISE} observed Phaethon at $\alpha$=25$^\circ$ (Table~\ref{obslog}), 
the $VLT$ was at $\alpha$=66$^\circ$ \citep[][]{Jewitt19Thermal}.   
The latter would see both the hot dayside and the cold nightside.  
For NEAs, the night-side flux could be important at high phase angle \citep{Harris1998Icar,Mommert2018AJ}, 
although this interpretation comes from classical thermal models 
with uncertain parameters (e.g. a beaming parameter: $\eta$, see \S~\ref{thermal}).
Regarding this, 2005~UD (MJD 58383: 2018-Sep-22) 
and 1999~YC (MDJ 57739: 2016-Dec16) taken at $\alpha >$\,75$^\circ$ by NEOWISE (Table~\ref{obslog}) 
would also include their night-side fluxes in the measurement.    
However our method is independent of those of models and the {\it WISE}/NEOWISE resolution is limited.       
Thus we assume 2005~UD and other NEOWISE data also have a blackbody-like temperature (Table~\ref{Teff}).

\subsection{Thermal Flux Density}
\label{thermal}

\begin{figure*}[htbp]
\epsscale{1} \plotone{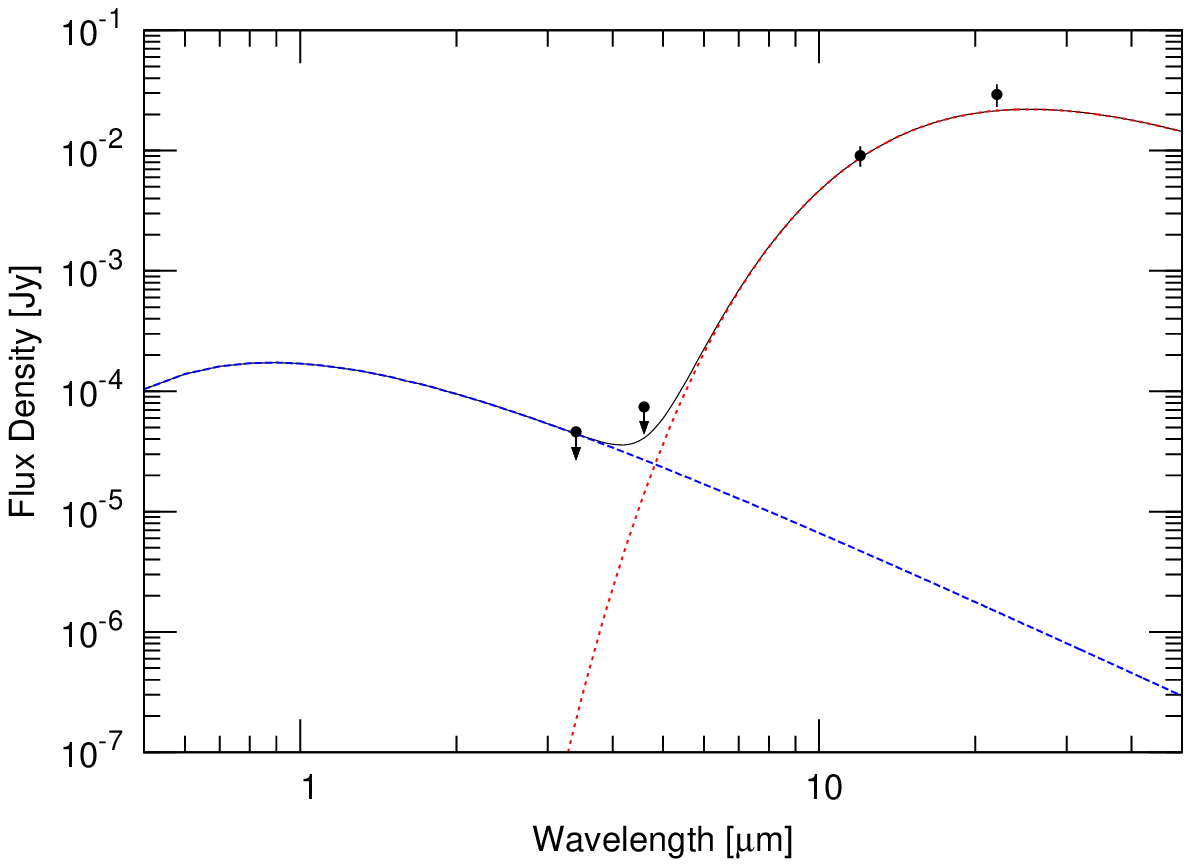} 
\caption{{\it WISE} observation of Phaethon on UT 2010 January 7 (MJD~55203.2564).
The measured flux densities at the $W1$~(3.4$\micron$), $W2$~(4.6$\micron$), 
$W3$~(12$\micron$) and $W4$~(22$\micron$) are shown as points.  
The 2$\sigma$ upper limit is represented with downward arrows.  
Reflected sunlight model (blue dashed-line), thermal model (red dashed-line) 
and combined signal (black solid-line) are over-plotted.   
}
\label{Phaethon_cntr1}
\end{figure*}

\begin{figure*}[htbp]
\epsscale{1} \plotone{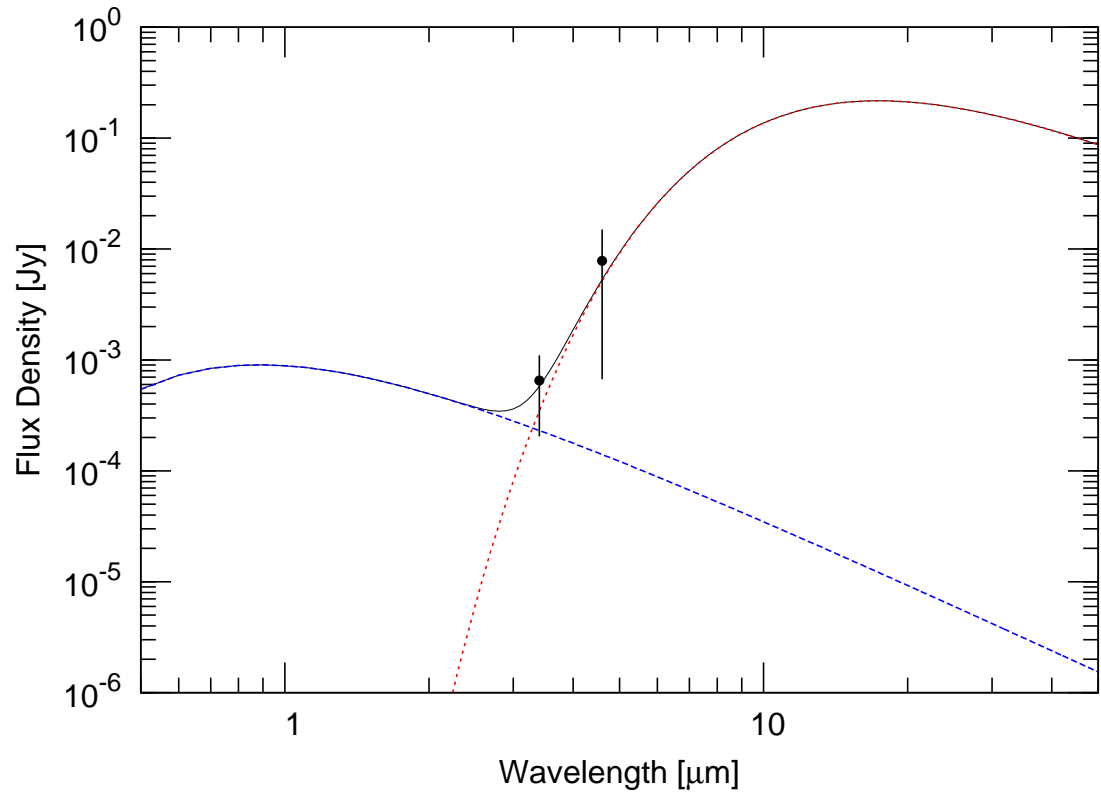} 
\caption{NEOWISE observation of 2005~UD on UT 2018 September 22 (MJD~58383.2977).  
Points show the measured flux densities at the $W1$~(3.4$\micron$) and $W2$~(4.6$\micron$).  
The model descriptions are same as Figure~\ref{Phaethon_cntr1}.  
}
\label{2005UD_cntr98}
\end{figure*}

\begin{figure*}[htbp]
\epsscale{1} \plotone{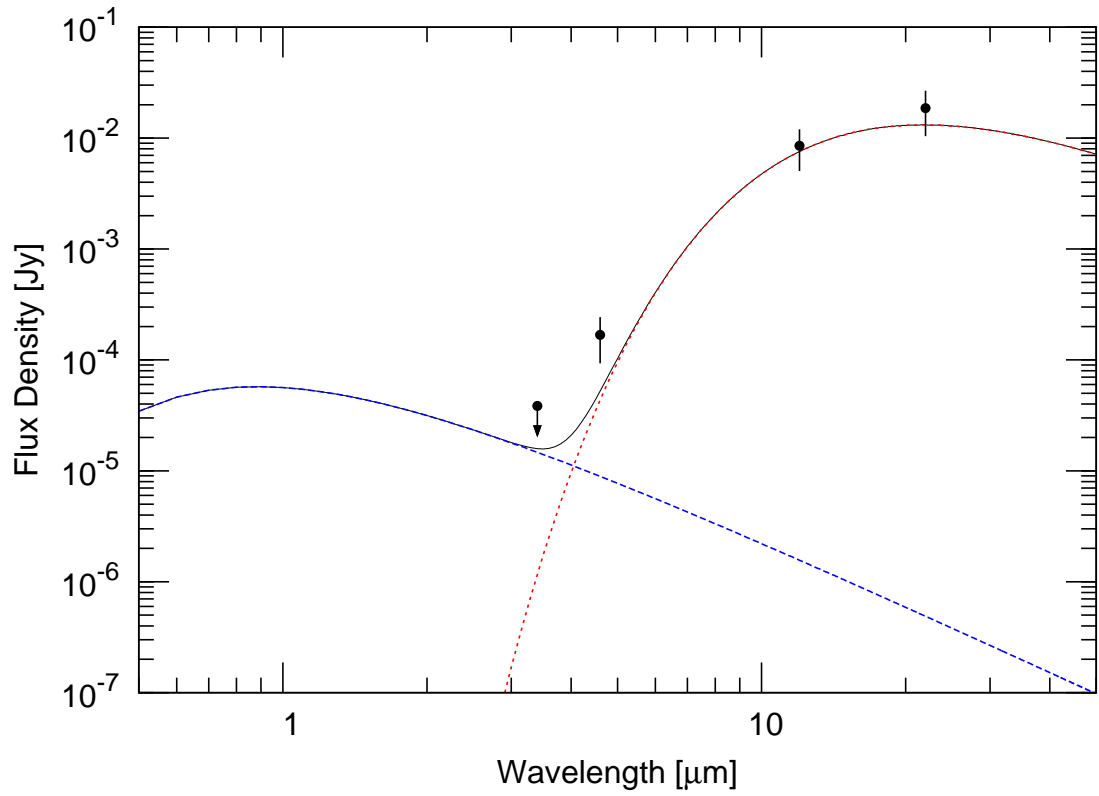} 
\caption{Same as Figure~\ref{Phaethon_cntr1} but for 1999~YC observed on UT 2010 January 10 (MJD~55213.6428).  
}
\label{1999YC_cntr5}
\end{figure*}

The thermal flux density from an asteroid, 
$F_\nu$~(Jy), 
is calculated using Equation~(1) of \cite{Mainzer2011ApJ}, 
\begin{equation}
F_\nu =  \epsilon \left(\frac{D_{\rm e}}{2}\right)^2 \frac{1}{\Delta^2} \int^{\pi/2}_{0} \int^{2\pi}_{0} B_\nu (T(\theta, \phi))\,d\phi\,{\rm sin}\,\theta\,{\rm cos}\,\theta\,d\theta,
\label{F_thermal}
\end{equation}
where $\epsilon$ = 0.9 is the infrared emissivity for the typical value 
of silicate powders from laboratory measurements \citep{HovisCallahan1966,Lebofsky1986Icar},
$B_\nu (T(\theta, \phi))$ is the Planck function (Jy~sr$^{-1}$) for the surface temperature $T(\theta, \phi)$ (K) 
in which $\theta$ (in degrees) is the angle from the sub-observer point to a point on the asteroid such that 
$\theta$ is equal to the phase angle $\alpha$ (in degrees) at the sub-solar point, and 
$\phi$ (in degrees) is an angle from the sub-observer point such that $\phi$ = 0 at the sub-solar point \citep[][]{Mainzer2011ApJ}.   
We again adopt the same values of $D_{\rm e}$ (cm) and $\Delta$ (cm) in section~\ref{Ref} (Tables~\ref{obslog} and \ref{Size_pv}). 
The surface temperature, $T(\theta, \phi)$, is calculated using Equation~(2) of \cite{Mainzer2011ApJ}, 
\begin{equation}
T(\theta, \phi) = T_{\rm ss}[{\rm max(0, cos\,\theta\,cos\,\alpha + sin\,\theta\,sin\,\alpha\,cos\,\phi)}]^{\frac{1}{4}},  
\label{Tss}
\end{equation}
where 
$T_{\rm ss}$ is the sub-solar temperature (K) derived by 
$T_{\rm ss}$ = $\chi^{\frac{1}{4}}$ $\cdot$ $T_{\rm eff}$ \citep{JewittLuu1992AJ}.  
The nondimensional parameter $\chi$ (1\,$\le$\,$\chi$\,$\le$\,4) represents 
the surface area ratio of the absorbing surface to the surface that is emitting the absorbed 
heat energy \citep{Jewitt_Kalas1998ApJ}.  
For instance, $\chi$=1 for a subsolar patch on a non-rotating asteroid 
and $\chi$=4 for a spherical, isothermal asteroid in which the Sun's heat is absorbed 
on $\pi$$r^2$ and radiated from 4$\pi$$r^2$ \citep{Li_Jewitt2015AJ}.
We adopt $\chi$ = 2 corresponding to thermal radiation from an asteroid hemisphere (observable-side).  
As examples,  
calculated spectral energy distributions and observed flux densities 
of Phaethon (MJD~55203.2564: 2010-Jan-07), 
2005~UD (MJD~58383.2977: 2018-Sep-22), 
and 1999~YC (MJD~55213.6428: 2010-Jan-10) 
is shown in Figure~\ref{Phaethon_cntr1}, \ref{2005UD_cntr98}, and \ref{1999YC_cntr5}, respectively.

We have a short note on the assumption of zero night-side emission.   
The classical thermal models generally use the nonphysical beaming parameter ($\eta$) 
for the sub-solar temperature ($T_{\rm ss}$) to adjust the angular temperature distribution (anisotropy) of the thermal emission,  
as seen in the Standard Thermal Model (STM), Fast Rotating Model (FRM) \citep[Reviewed in][]{LebofskySpencer1989aste} 
and the near-Earth asteroid thermal model (NEATM) \citep{Harris1998Icar}. 
The common issue in those of models is that the strong dependence on $\eta$ 
which is not an intrinsic property for the temperature distribution over the observable surface of asteroid.    
Also it varies with observing geometry ($\alpha$, $R_{\rm h}$ and $\Delta$), as well as 
the physical properties (e.g. thermal inertia) of the asteroid \citep[See \S 3.4, ][]{Wright2018arXiv} \citep[See also][]{Masiero2019AJ}. 
Thus we avoid the ambiguity of $\eta$.

\section{Results}

Our images of Phaethon, 2005~UD and 1999~YC each 
look like point sources and show no apparent evidence of coma activity.   
When both the reflected sunlight and thermal emission 
from the asteroids are removed (\S~\ref{Ref} and \ref{thermal}), 
residuals remain in some data.  
For active comets, 
the excess signals in the $W1$ and $W2$ bands of {\it WISE}/NEOWISE 
can be used to estimate the dust and gas (CO and CO$_2$) 
production rates respectively \citep{Bauer2008PASP,Bauer2011ApJ,Bauer2015ApJ,Bauer2021PSJ,Mainzer2014ApJ}.  
The longer wavelength bands ($W3$ and $W4$) are capable of 
examining the dust trails and potentially detect co-moving objects \citep[cf.][]{RSL2000,Reach07,Jewitt19Thermal}.  
Here, we asses the significance of the excess signals and attempt to 
set an empirical upper limit to the production rates, dust trails and co-moving objects.

\subsection{Dust Production Rate: $W1$}
\label{W1Q}

We look into the dust production rate with the $W1$ (3.4\,$\micron$) data 
following the methodology in \cite{Bauer2008PASP,Bauer2011ApJ,Bauer2021PSJ}.  
The reflected dust signal is typically considered to dominate in the band for all but the nearest NEOs.   
Even when comets sublimate their abundant icy volatiles,  
dust comprise $\gtrsim$~70\% of the total signal 
\citep{Bockel1995Icar,Reach2013Icar} \citep[Summarized in][]{Bauer2015ApJ}.  
Assuming the excess flux density is produced by coma brightness,   
the mass of dust particles scattering in the coma, 
$M_d$ (kg), 
is given by \cite[Equations~(3) and (5) of][]{Bauer2008PASP} 
\begin{equation}
M_d = \frac{a_d\,\rho_d}{3} \times \pi D_e^2~\frac{F_{{\rm total}} - F_{\rm calc}}{F_{\rm calc}}, 
\label{mass}
\end{equation}
where 
$a_d$ = 1.7\,$\micron$ \citep[1/2 of the $W1$ wavelength,][]{Bauer2011ApJ} 
is the radius of the dust particle contributing to coma brightness,  
$\rho_d$ = 2000~kg\,m$^{-3}$ is the assumed common 
bulk density of dust and asteroid \citep[cf. Phaethon,][]{Hanus2018AA}, 
$D_e$ (m) is the diameter of the asteroid, 
$F_{\rm total}$ (Jy) is the measured total flux density from an asteroid, 
and 
$F_{\rm calc}$ (Jy) is the calculated flux density 
from an asteroid (= reflected sunlight + thermal emission).   
The dust production rate, $Q_{\rm dust}$ (kg s$^{-1}$), 
is estimated using $M_d$ (kg), 
velocity of dust, $v_d$ (km s$^{-1}$), 
and projected size of the dust coma, $R_d$ (km) \citep[$\approx \rho$ in the $Af\rho$ method,][]{AHearn1984}.
The traveling time for the produced dust from the asteroid is   
$t_d$ (second) = $R_d/v_{d}$, and using the Equation~(6) of \cite{Bauer2008PASP}, we obtain the relation  
\begin{equation}
Q_{\rm dust} = \frac{M_d}{t_d} = \frac{M_d}{R_d} v_d,    
\label{Qdust}
\end{equation}
where 
$R_d$\,(km) is the size of subtended area using an aperture radius of the of 7$\arcsec$.5 at the ${\it WISE}$-centric distance $\Delta$ (au), 
and 
$v_{d}$ 
is adopted as the escape velocity from the PGC asteroids, 
i.e. $v_{d}$ $\sim$ 3\,m\,s$^{-1}$ for Phaethon and $v_{d}$ $\sim$ 1\,m\,s$^{-1}$ for 2005~UD and 1999~YC, respectively.   
The obtained mass loss rates are shown in Table~\ref{W1}. 

We find no strong variation in the production rate at different epochs,
and find no dependency on the location of the object in the orbit at the time of observation.  
The upper limit to the mass-loss 
rate for Phaethon is $Q_{\rm dust}$ $\lesssim$ 2\,kg\,s$^{-1}$ and 
those of 2005~UD and 1999~YC are 
both $Q_{\rm dust}$ $\lesssim$ 0.1\,kg\,s$^{-1}$, respectively.   
The {\it WISE}/NEOWISE-derived rate limits 
are 1--2 orders of magnitude larger than 
those of optically measured limits at the $R_c$-band 
of 0.001 -- 0.01\,kg\,s$^{-1}$ 
\citep{HsiehJewitt2005,JewittHsieh2006,KasugaJewitt2008AJ}.  
Meanwhile, our $Q_{\rm dust}$ is $\sim$10 times lower than 
the limit of $\lesssim$~14\,kg\,s$^{-1}$ measured 
at the mid-infrared band ($\lambda$=10.7$\micron$) from Phaethon \citep{Jewitt19Thermal}.

The estimated limiting mass-loss rates of Phaethon can be 
compared with the total mass of the Geminid meteoroid stream. 
The Geminid stream mass, 
$M_s$ $\approx$ $10^{13}$\,kg  -- $10^{14}$\,kg \citep[][]{Bla17} 
is thought to have been produced from Phaethon over the last $\tau_s \sim 1000(s)$ years 
\citep[e.g.][]{WW93}.    
The steady-state mass-loss rate needed, 
$dM_s/dt \sim M_s/\tau_s$ $\sim$ 320 -- 3200~kg\,s$^{-1}$, 
is comparable to those of active Jupiter family comets (JFC). 
This requires at least 10--100 times larger rates than 
the observed rates in the near- and mid-infrared bands (3.4$\micron$ and 10.7$\micron$).   
The leading conclusion is that the production of the Geminids occurred episodically, 
such as catastrophic breakup of a precursor body, 
not in steady disintegration at the observed rates \citep[cf.][]{Jewitt19Thermal}. 
Likewise, to estimate the mass-loss rate ($Q_{\rm dust}$) from 2005~UD, 
we estimate the mass of the Daytime Sextantid meteoroid stream.  
The stream age is calculated to be $>$10$^4$ yr \citep{JakubNeslu2015}, 
comparable to the last lowest perihelion-passage \citep[$\approx$\,2$\times$10$^4$ yr ago,][]{MacLennan2021Icar}.    
Assuming that the meteoroids have been 
released from 2005~UD over the age of 10,000\,yr,  
the maximum mass-loss rate $Q_{\rm dust}$ $\lesssim$ 0.1\,kg\,s$^{-1}$ 
finds the total mass of the Daytime Sextantid stream $\sim10^{10}$\,kg.  
Note that if 2005~UD shared a precursor body with Phaethon, 
the resulting breakup event could provide more mass in the stream. 
To investigate the possibility of a breakup-induced stream formation,
longer wavelength observations of the parental asteroids
to detect larger particles, potentially containing much more 
mass, are worthwhile to continue \citep{Jewitt_active15,KasugaJewitt2019}.    
Further observations of the Geminids and the Daytime Sextantids and 
connection to a stream model 
will be helpful to understand the stream formation processes.

\subsection{Gas (CO and CO$_2$) Production Rate: $W2$}
\label{W2Q}

Active comets showing excess flux density  
in the $W2$ (4.6$\micron$)-band is attributable to gas emission lines, in particular 
the CO$_2$ $\nu_3$ vibrational fundamental band (4.26$\micron$) 
and the CO $v(1-0)$ rovibrational fundamental bands (4.67$\micron$) 
\citep[][]{Pittichova2008AJ,Ootsubo2012ApJ,Bauer2011ApJ, Bauer2015ApJ,Bauer2021PSJ,Rosser2018AJ}. 
These volatiles have low sublimation temperatures of 20--100\,K, 
presumed to be preserved in frozen ices or trapped as gases 
in the nuclei \citep{Prialnik2004come.book,BouzianiJewitt2022}.  

This gas emission is unlikely to be present in the case of the PGC-asteroids 
because of their thermophysical and dynamical properties \citep{Jewitt18HST,Jewitt19Thermal}.
The surface temperatures at perihelia are $T_{\rm ss}^{\rm PGC}$~$\gtrsim$~800\,K, too hot for ice to survive.   
The largest plausible thermal diffusivity $\kappa \sim 10^{-6}\,{\rm m^2\,s^{-1}}$ is 
appropriate for rock (a compact dielectric solid).  
The diurnal thermal skin depth, 
$d_{\rm s}$, is estimated by $\sim$ $\sqrt{\kappa P_{\rm rot}}$, where $P_{\rm rot}$ is the rotational period.  
Setting $\kappa$ = 10$^{-6}$\,${\rm m^2\,s^{-1}}$ and 
$P_{\rm rot}$ = 3.6 -- 5.2\,hr for the PGC asteroids \citep[Table~2,][]{KasugaJewitt2019}
find $d_{\rm s}$~$\sim$~0.14\,m at deepest.  
The $q$-temperature at the thermal skin depth is $>$~200\,K (= $T_{\rm bb}^{\rm PGC}$/$e$), 
far above the sublimation temperature of CO$_2$ and CO.  
The heat conduction timescale corresponding to the equator radius of Phaethon ($r \sim$ 3\,km) 
is 0.3\,Myr ($\approx$ $r^2$/$\kappa$),  
about two orders of magnitude shorter than its dynamical lifetime of 26\,Myr \citep{deLeon10}.
The core temperatures of the PGC asteroids are sufficiently heated at 
$T_{\rm core} \gtrsim$ 280\,K \citep[Equation~(4) of][]{JewittHsieh2006}.  
Dynamical simulations find that Phaethon on its present orbit 
could lose all internal ice (e.g. H$_2$O) over a very short timescale of 5--6\,Myr \citep{Yu2019,MacLennan2021Icar}.  
Therefore, the PGC-asteroids are highly unlikely to be reservoirs for 
the icy species (CO$_2$ and CO) which are not expected to survive 
on the surfaces or in the interiors.   
We thus do not make further discussion on this topic.

\subsection{Dust Trail: $W3$ and $W4$}
\label{dusttrail}

We search for dust trails that presumably formed from the mass-loss of the PGC-asteroids.  
Such trails consist of larger, slow-moving dust particles 
that would be insensitive to solar radiation pressure.  
These particles would follow the parent bodies' heliocentric orbits and stay close to those 
orbital planes \citep[Reviewed in][]{Jewitt_active15}.   
Previous thermal observations ({\it DIRBE}/{\it COBE}) managed to detect the surface brightness of 
the Phaethon's dust trail, consisting of larger particles with sizes $>$10~$\micron$ \citep{Arendt14}.
For all of the small bodies observed by ${\it WISE}$/NEOWISE,  
the both $W3$- and $W4$-bands are dominated by thermal emission \citep{Mainzer2011ApJ}.  
Since the PGC-asteroids have blackbody-like temperatures (Table~\ref{Teff}), 
the Wien's displacement law 
($\lambda_{\rm max}$~$\cdot$~$T_{\rm eff}$ = 2898\,\micron\,K gives 
$\lambda_{\rm max}$=10.7$\sim$15.8$\micron$) indicates that 
the $W3$-band (12\,$\micron$) is the most effective wavelength for larger particles.  
Thus we primarily study the $W3$-data of Phaethon and 1999~YC, 
and the $W4$-data is supplementary compared.        
Unfortunately {\it WISE} did not observe 2005~UD during the cryogenic mission, so we do not have $W3$ or $W4$ data for this object.        

During the survey, the apparent motion of the asteroids with respect to 
the {\it WISE} location had rates up to approximately 28$\arcsec$~hr$^{-1}$ 
in right ascension (RA) and $-$32$\arcsec$~hr$^{-1}$ 
in declination (DEC)\footnote{NASA JPL HORIZONS \url{https://ssd.jpl.nasa.gov/horizons/}}.  
The corresponding drift across the frame is $\lesssim$ 0.1$\arcsec$ during each exposure.
These rates are more than an order of magnitude smaller 
than the pixel scale (2\arcsec.76 pix$^{-1}$), 
meaning that smearing within an image is negligibly small. 
For this analysis, we do not include data in which asteroid is taken 
close to edge of the frame, where trail location is expected 
to be out of FOV.   Also we avoid images in which 
trail's expected direction is contaminated by stars.       
Detectability of a trail is improved for small out-of-plane angles, $\delta_\Earth$, 
which is given by the angle between the observer and target orbital planes \citep{Jewitt18HST}. 
The $\delta_\Earth$=3.8$^\circ$ for Phaethon indicates 
it should be highly sensitivity to the dust trail close to the plane, 
while $\delta_\Earth$=$-$23$^\circ$ for 1999~YC is slightly worse.  

The $W3$ image was corrected by subtracting a bias 
constructed from a median-combination of applied images (Table~\ref{obslog}).  
The orientation of each image was rotated to bring 
the direction of the position angle (PA) of north to the top 
and east to the left, and shifted to align the images using 
fifth-order polynomial interpolation. The images were then combined into a single summed image.
The resulting summed image of Phaethon has a FWHM of $7\arcsec.8$, 
and is shown in Figure~(\ref{PhaethonW3}).  
Likewise, the summed image of 1999~YC with a FWHM of $7\arcsec.1$ 
is shown in Figure~(\ref{1999YCW3}).  
No asteroid-associated dust trails, which are expected to be 
parallel to the $``-V"$ vector, are apparent in either Figure.

In general, highly-resolved observations of 
asteroidal dust trails have clarified their morphological properties.   
For example, the trail widths of active main-belt 
asteroids are commonly very narrow, near the 
bodies, and represent rather recently ejected debris.  
The FWHM of trail widths of 300--600\,km ($<$~1$\arcsec$) are typically very narrow, 
as measured from 133P/Elst-Pizarro and 311P/PanSTARRS (P/2013 P5) \citep[\S 3.1,][]{Jewitt18HST}.  
Small ejection velocities of $\sim$ 1--2\,m\,s$^{-1}$ in the perpendicular direction to the orbital plane, 
comparable to the gravitational escape speed ($V_{\rm esc}$), are measured for most active asteroids \citep[][]{Jewitt_active15}.
    
However, this expectation is unlikely to hold for Phaethon.  
The width of the Phaethon dust trail (Geminid stream) is expected to be 
wider than those of asteroidal trails. 
Optical observations of the Geminid meteor shower estimate the shower duration time.  
The shape of zenithal hourly rate (ZHR) of the Geminids is asymmetric at the peak time 
with its minimum FWHM of $\gtrsim$~1$^\circ$ in the solar longitude, 
corresponding to the duration time $\gtrsim$~24\,hr \citep{Uchiyama2010JIMO}.
Geometrically, the Earth ($v_\Earth$=30\,km\,s$^{-1}$) cuts through the stream 
at the crossing angle of 62$^\circ$ on the projected ecliptic plane (Mikiya Sato, private communication).   
Thus we find the Geminid stream width has a FWHM $\sim$ 2.3$\times$10$^{6}$\,km    
(= 30\,km\,s$^{-1}$ $\times$ 24\,hr $\times$ 3600\,s\,hr$^{-1}$ $\times$ sin(62$^\circ$)).    
For 1999~YC, neither its activity nor trail has been observed yet \citep{KasugaJewitt2008AJ}.    
Assuming that 1999~YC had a trail, and the width FWHM would be 
at most $V_{\rm esc}$$\cdot$$P_{\rm orb}$/4, 
where $P_{\rm orb}$ is the orbital period and 4 is because a particle 
nearly spends 1/4 orbit rising from the orbital plane to the peak height 
(David Jewitt, private communication).   
Substituting $V_{\rm esc}$$\sim$1\,m\,s$^{-1}$ and $P_{\rm orb}$=1.7\,yr, 
the trail width FWHM would be expected to be about 1.3$\times$10$^{4}$\,km. 
Such a trail could be imaged by {\it WISE}, 
if it existed and was bright enough, as the WISE images have a resolution of 2,400\,km\,pix$^{-1}$ 
at the observed $\Delta$=1.22\,au to 1999~YC.

To measure surface brightness profiles 
perpendicular to the expected trail direction (``$-V$"), 
the summed images of Phaethon and 1999~YC were each 
rotated to set the projected orbit direction to the horizontal.
Both the $W3$ and $W4$ images are used for comparison.    
We average $\pm$1~pix (2.\arcsec76 for $W3$ or 5.\arcsec5 for $W4$) 
left and right of the object at a distance of 50--215$\arcsec$ from the asteroids.  
The profile for Phaethon and 1999~YC is shown in 
Figure~(\ref{PhaethonW3W4}) and (\ref{1999YCW3W4}), respectively.  
The green region corresponds to approximate width of trail.  
For Phaethon (Figure~\ref{PhaethonW3W4}), 
a trail would be detected as a symmetric excess at 
x = 0\arcsec and $\sim$1530\arcsec wide if FWHM $\sim$ 2.3$\times$10$^{6}$\,km.
Likewise for 1999~YC (Figure~\ref{1999YCW3W4}), 
the calculated trail width FWHM $\sim$ 1.3$\times$10$^{4}$\,km 
has a symmetric green region centered at x = 0\arcsec and $\sim$16\arcsec wide.  
No excess is evident in the profiles shown, 
neither in the $W3$-band nor in the $W4$-band 
measured at distinctive distance combinations of vertical cuts.   
We find no any hint of an orbit-aligned trail, or any thickness.   

Here, we have a short note on 
the non-detection of the Phathon trail by ${\it WISE}$.  
This result is contrary to {\it DIRBE}/{\it COBE}, probably 
caused by observing geometry.  
{\it WISE} measured Phaethon and its vicinity 
at the solar elongation $\varepsilon \sim$91$^\circ$ and $R_{\rm}$=2.32\,au.   
On the other hand, 
{\it DIRBE}/{\it COBE} detected the surface brightness of the trail  
at $\varepsilon \sim$64$^\circ$ and $R_{\rm}$=1.01\,au  \citep{Arendt14}, 
when it was much closer to the sun, so much warmer and less dispersed.  
{\it WISE} was at a rather disadvantageous observing point for trail thermal emission.  
Despite this, the background in the {\it WISE} images 
can set a uniformly applicable limit for constraining signals referring 
the morphological evidence from the asteroids.

We set a practical upper limit to 
the surface brightness of the potential dust trails.  
The $W3$-band (12$\micron$) is inspected because of the most efficient SNR.  
We sampled 70\,pix (193$\arcsec$) along the trail direction from the each asteroid.  
The profile was averaged along the rows over the width FWHM of the trail.  
Two-consecutive highest counts were selected and  
averaged between the distances for placing a statistical limit.   
For the Phaethon trail, we used the angular distances 
33\arcsec$\leqslant \theta_{\rm od} \leqslant$36\arcsec which 
corresponds to the distances of 5.0$\sim$5.4$\times$10$^{4}$\,km from Phaethon.
For the 1999~YC trail, we used the angular distances of 
69\arcsec$\leqslant \theta_{\rm od} \leqslant$72\arcsec which  
corresponds to the distances of 6.1$\sim$6.4$\times$10$^{4}$\,km from 1999~YC.
The measured counts (in digital numbers) were converted to the flux density (in Jy, see \S~\ref{magflux}).  
The ratio of 
the measured flux density scattered by the dust particles in the trail ${\it I_d}$ (Jy) 
to the measured flux density scattered by the asteroid (nucleus) cross section ${\it I_n}$ (Jy),
which corresponds to the ratio of 
the dust cross section in the trail ${\it C_d}$ (km$^2$) 
to the asteroid (nucleus) cross section ${\it C_n}$ (km$^2$), 
can be calculated \citep{Luu1992Icar}, and is expressed as 
\begin{equation}
\frac{I_d}{I_n} = \frac{C_d}{C_n}.  
\label{ratio}
\end{equation}
The optical depth was obtained from $\tau$ = ${\it C_d}$/$s^2$, 
where $s$(km)=725.27$\cdot$$\Delta$ 
is the linear distance (corresponding to 1 arcsecond)   
at the ${\it WISE}$-centric distance $\Delta$(in au) of the asteroid.   
The results are summarized in Table~\ref{optdepth}.

The upper limit ($> 3$$\sigma$) to optical depth of the dust trail is 
$\tau$~$<$~7$\times$10$^{-9}$ for Phaethon and 
$<$~2$\times$10$^{-8}$ for 1999~YC, respectively. 
These are broadly comparable to the optical depths of 
cometary trails (10$^{-10}$--10$^{-8}$) measured by 
thermal infrared surveys \citep{SykesWalker92,RSL2000,Reach07}.  
These results are commonly obtained at far distance away ($>$~10$^4$\,km) 
from the asteroids or nuclei.   
At much closer distances of 150$\sim$200\,km for Phaethon, on the other hand, 
we would estimate optical depths three orders of magnitude higher: $\tau$~$<$~6$\times$10$^{-6}$ 
\citep[][]{Jewitt19Thermal}. 
This implies that any larger dust particles ($>$10$\micron$) are  
present near the parent body.    
     
We note that optical depths measured at similar wavelengths 
can be directly compared, because of the similar particle size 
dependence of the radiating efficiency \citep[][]{Jewitt19Thermal}.  
While thermal observations ($\lambda$~$>$~10$\micron$) for  
cometary dust trails have obtained optical depths, 
those of asteroidal dust trails (bands) \citep[e.g.][]{Nesvorny2006Icar,Nesvorny2006AJ,Espy2015} 
and bolides \citep{Borovicka2020AA} are mostly undetermined.   
More samples would be helpful for understanding 
optical depths of asteroidal dust trails.

\subsection{Size Limits for Co-moving Objects: $W3$}

\begin{figure}[ht!]
\epsscale{1}\plotone{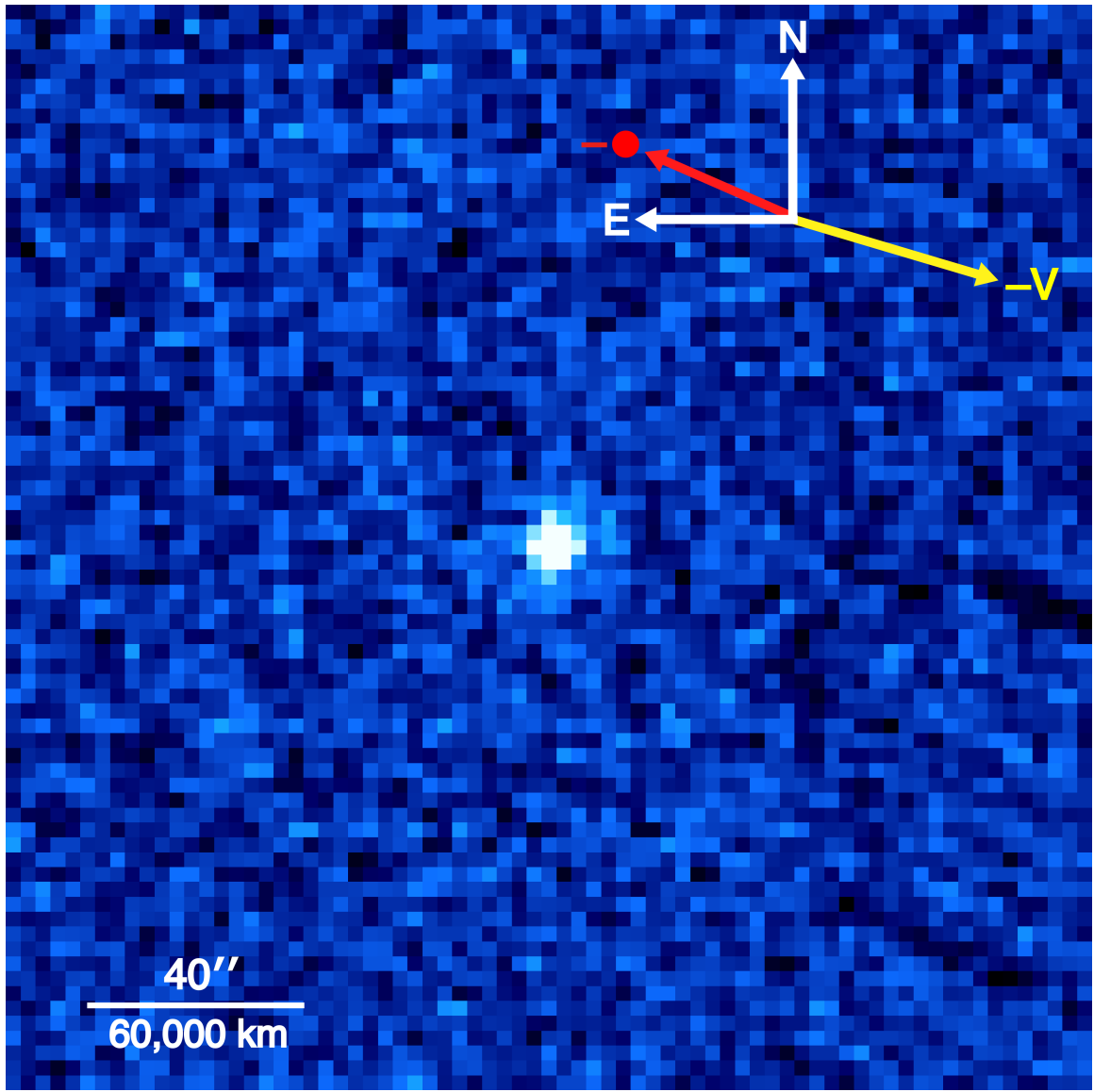}
\caption{Composite $W3$-band image of Phaethon in 17.6\,s integration (2~$\times$~8.8\,s) taken by {\it WISE} on UT 2010 January 7.   
The frame size is 200$^{^{\prime \prime }}$ $\times$ 200$^{^{\prime \prime }}$. 
No coma or tail is visible on the object having an FWHM of 7.8\arcsec.  
Heliocentric, ${\it WISE}$-centric distances and phase angle were $R_{\rm h}$=2.32\,AU, $\Delta$=2.08\,AU and $\alpha$=25.1$^{\circ}$, respectively. 
The cardinal directions (N and E), the direction of the negative heliocentric velocity vector ($-V$), 
and the anti-solar direction ($-\bullet$) are marked.  
A 40$\arcsec$ scale bar corresponding to 60,000\,km at the $\Delta$ distance to Phaethon is shown.
} 
\label{PhaethonW3}
\end{figure}

\begin{figure}[ht!]
\epsscale{1}\plotone{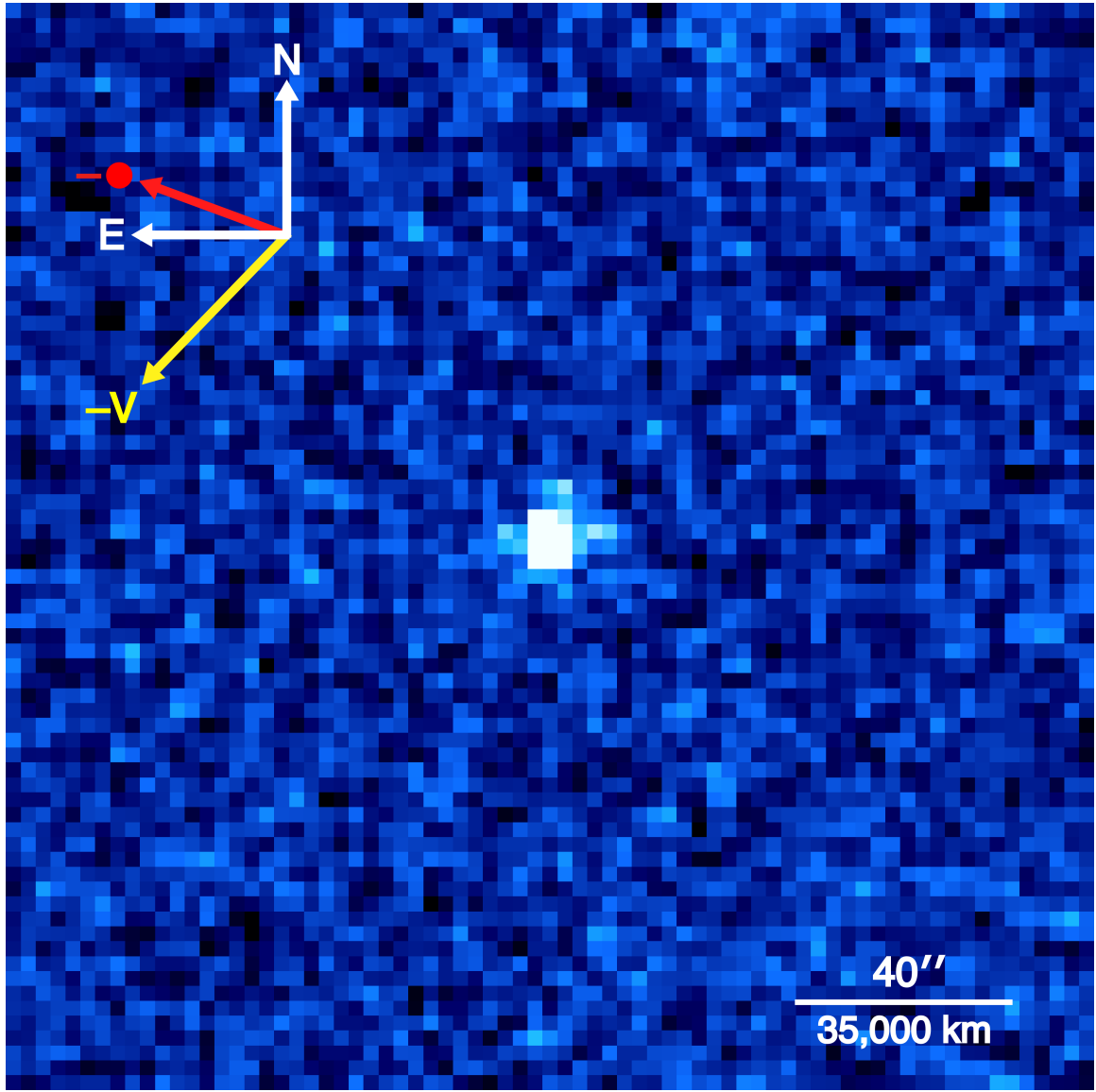}
\caption{${\it WISE}$ $W3$-band image of 1999~YC on UT 2010 January 10 
showing a point source with FWHM$\sim$7.1\arcsec~in 35.2\,s integration (4~$\times$~8.8\,s), 
centered in the frame of 200$^{^{\prime \prime }}$ $\times$ 200$^{^{\prime \prime }}$.  
$R_{\rm h}$=1.62\,AU, $\Delta$=1.22\,AU and $\alpha$=37.3$^{\circ}$.  
$N$ and $E$ exhibit the cardinal directions.  
$-V$ shows the direction of the negative heliocentric velocity vector 
and 
$-\bullet$ shows the anti-solar direction, respectively.  
A 40$\arcsec$ scale bar corresponding to 35,000\,km at the $\Delta$ distance to 1999~YC is shown.
}
\label{1999YCW3}
\end{figure}

\begin{figure}[ht!]
\epsscale{1}\plotone{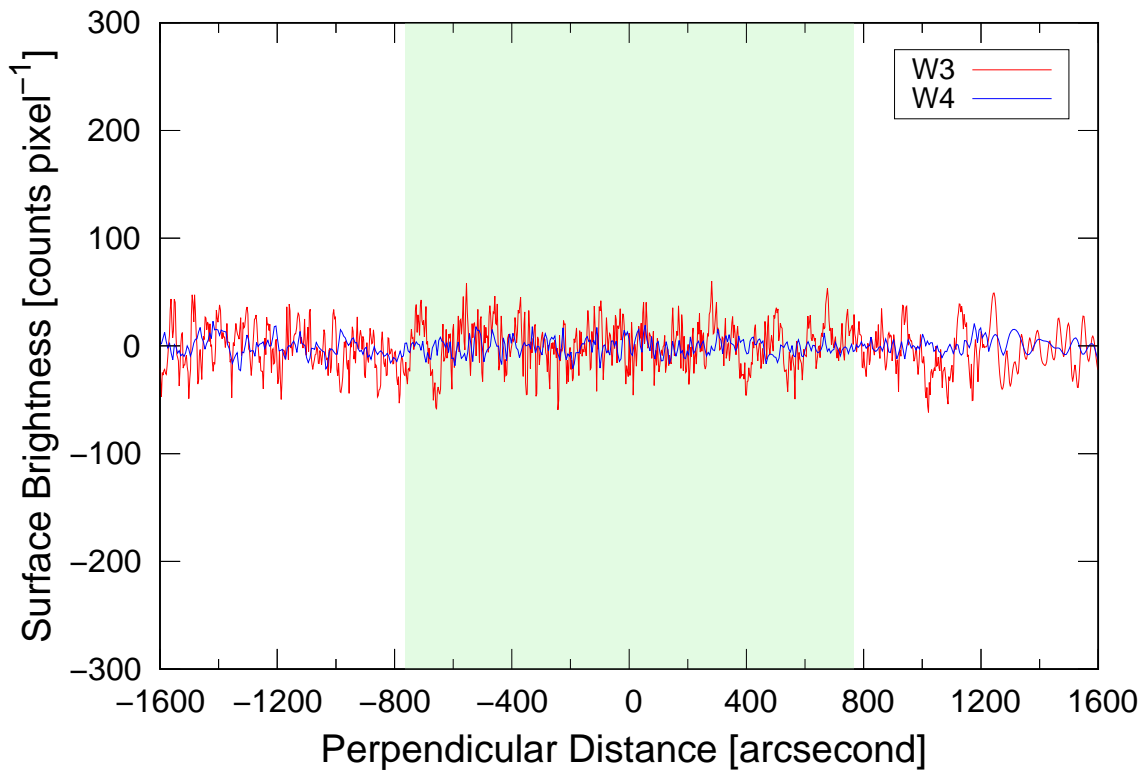}
\caption{Averaged surface brightness cuts perpendicular to the projected orbital plane of Phaethon.  
Red and blue line shows a random cut measured in the $W3$- and $W4$-band at the expected trail location ($-V$), respectively.  
The $W3$-band profile was obtained at distance 50\arcsec ($\approx$75,000\,km) from Phaethon while, 
in the $W4$-band, the profile was extracted at distance 72\arcsec ($\approx$110,000\,km).   
The green region shows the expected location and approximate width of the in-plane dust trail.  
Surface brightness expressed in digital units (1\,count per pixel) corresponds to 20.0\,mag\,arcsec$^{-2}$ at $W3$ 
and 16.6\,mag\,arcsec$^{-2}$ at $W4$.  
}
\label{PhaethonW3W4}
\end{figure}

\begin{figure}[ht!]
\epsscale{1}\plotone{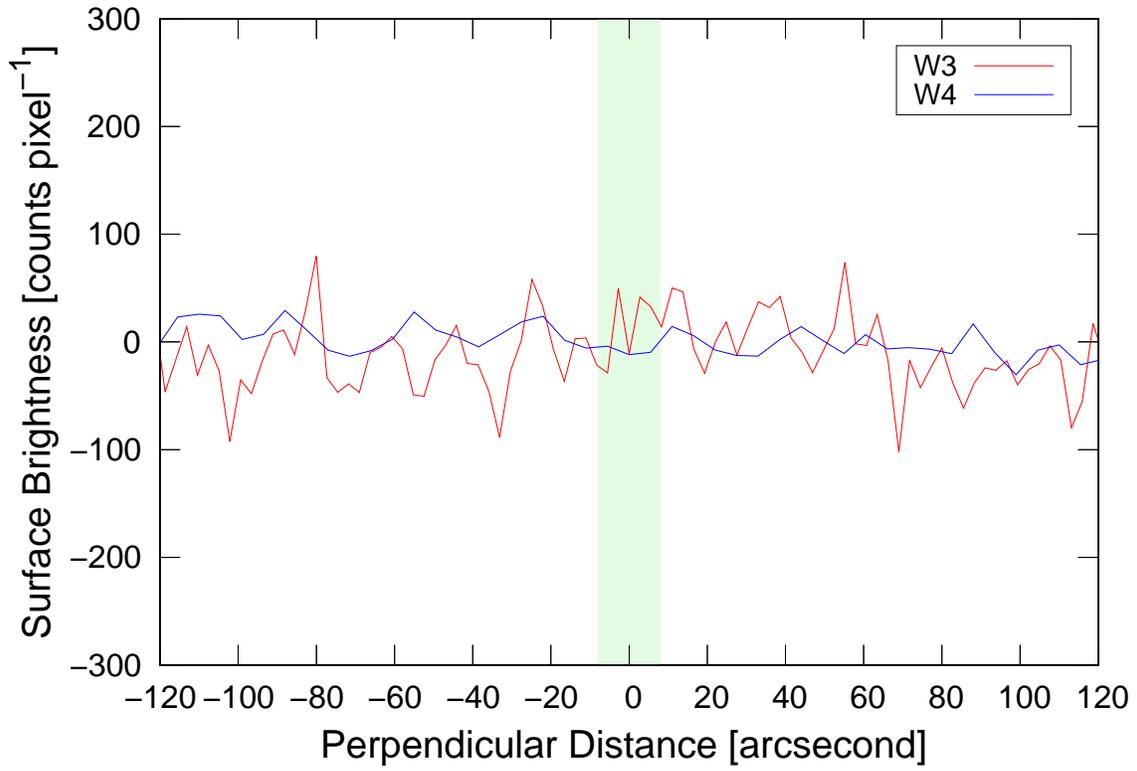}
\caption{Same as Figure~\ref{PhaethonW3W4} but for 1999~YC.  
The $W3$-band profile is at distance 160$\arcsec$ ($\approx$140,000\,km) from 1999~YC 
and the $W4$-band profile is at distance 215$\arcsec$ ($\approx$190,000\,km). 
}
\label{1999YCW3W4}
\end{figure}

We searched for possible companions around the PGC-asteroids.   
Again, the $W3$-images ({\it WISE}) were used. 
As in the Figures~\ref{PhaethonW3} and \ref{1999YCW3}, 
no co-moving objects are apparent in our data.
The kilometer-sized asteroids would be immediately 
obvious, while smaller bodies could linger and might just escape detection. 
Here we set limits to the brightness of possible co-moving point sources.  
We put an aperture down in blank regions and measured 
flux counts and the uncertainty in digital numbers.  
The aperture area was set to 100\,pix$^2$ (10\,pix$\times$10\,pix), comparable to those of the PGC asteroids.   
The projected angular distance was set at $\sim$28\arcsec (10\,pix), 56\arcsec (20\,pix) 
and 84\arcsec (30\,pix)~in the North direction from each asteroid (Figures~\ref{PhaethonW3} and \ref{1999YCW3}).  
The corresponding shortest distance is 4.2$\times$10$^{4}$\,km from Phaethon 
and 2.4$\times$10$^{4}$\,km from 1999~YC, respectively, which corresponds to 
far outside of their Hill radii ($\lesssim$\,60\,km~$\approx$~0.04\arcsec).   
Since no distance dependency is found in the measured counts,  
we applied those of the averaged value to obtain flux density, $I_{\rm e}$ (in Jy).  
In order to place the limits to size, we scale them from 
Phaethon and 1999~YC (see $I_{\rm n}$ in Table~\ref{optdepth}, respectively) 
by assuming that the flux density is proportional only to 
the cross-sectional area of the radiating body. 
The derived flux density ratio, $I_{\rm e}/I_{\rm n}$, 
is 5.8$\pm$0.7$\times$10$^{-2}$ in Phaethon 
and 2.8$\pm$0.4$\times$10$^{-2}$ in 1999~YC, respectively.
The limiting radius of possible companions, $r_{\rm e}$~(m), is given by 
$r_{\rm e}$= 1000\,$(D_{\rm e} /2)$\,$(I_{\rm e}/I_{\rm n})^{0.5}$.  
Substituting $D_{\rm e}$ (Table~\ref{Size_pv}) and $I_{\rm e}/I_{\rm n}$, we obtain 
the 3$\sigma$ upper limit to radius of possible co-moving object, $r_{\rm e}$ $<$ 650\,m 
in Figure~\ref{PhaethonW3} (Phaethon) 
and $r_{\rm e}$ $<$ 170\,m in Figure~\ref{1999YCW3} (1999~YC), respectively.
Objects down to the size limits would be apparent in blank 
space of the $W3$-images though, undetected.

\section{Discussion}

\subsection{DESTINY$^+$ Update}

JAXA's DESTINY$^{+}$ mission plans a Phaethon flyby 
at a relative speed of 36\,km\,s$^{-1}$ 
either in 2028 January or 2030 November, 
with a closest approach distance of about 500\,km 
\citep{Ozaki2022arXiv}.  
The DESTINY$^{+}$ dust analyzer (DDA) has 
the two sensor heads with a total sensitive 
area, $A_{\rm DDA}$= 0.035\,m$^2$, for collecting 
interplanetary dust particles (IPDs) with 
radii $\lesssim~10\micron$ (= a particle mass $\lesssim$\,10$^{-11}$\,kg) \citep{Kruger2019PSS}.   
But {\it HST} optical observation reported that 
the DDA is insufficient for sampling the Geminid meteoroids 
during the Phaethon flyby \citep{Jewitt18HST}.
Here we have an update with the thermal observations 
conducted at the two distinctive distances from Phaethon.   
One measurement was at the short distance of 150$\sim$200\,km to Phaethon \citep[][]{Jewitt19Thermal} 
and another was at the far distances of 50,000$\sim$54,000\,km based on this {\it WISE} study.  
We follow the procedure described in \cite{Jewitt18HST}.  

We examined the DESTINY$^{+}$ main-unit and DDA, respectively.  
Here, we describe the main-unit.   
The DESTINY$^{+}$ is a cuboid shape having a longitudinal cross section $A_{\rm D^+}$ 
$<$ 1.7\,m$^2$(Naoya Ozaki, private communication)
~\footnote{See \url{https://jpn.nec.com/press/202101/20210128_02.html} in Japanese}. 
The total cross section of all the particles to be intercepted in this area is $C = \tau A_{\rm D^+} $.
This is equal to the maximum cross section of a single, spherical particle having radius $a <$ $(\tau A_{\rm D^+} /\pi)^{1/2}$.  
The corresponding particle mass is $M=4/3 \pi \rho_d a^3$, where the assumed bulk density $\rho_d$=2000\,kg\,m$^{-3}$.
An upper limit to the number density of dust particles, $N_1$ (m$^{-3}$), 
is placed using either the measured optical depth ($\tau$) as expressed by $N_1 =  \tau / (\pi a^2 L$),  
or the measured mass-loss rate ($Q_{\rm dust}$) as expressed by \citep[Equation~(3) of ][]{Jewitt18HST}, 
\begin{equation}
N_1 = \frac{1}{4 \pi U L^2}\left(\frac{3}{4 \pi \rho_{d}a^3}\right)Q_{\rm dust},
\label{N1}
\end{equation}
where $a$ is the maximum radius of dust particles, 
$L$ is the path length along the line of sight, 
and $U$ is the speed of released dust particles from Phaethon. 
We apply $L$=500\,km for the closest distance of DESTINY$^{+}$ 
with the measured values of $\tau$\,$<$\,6$\times$10$^{-6}$ 
and $Q_{\rm dust}$\,$\lesssim$\,14\,kg\,s$^{-1}$ \citep[][]{Jewitt19Thermal}
\footnote{We assessed a linear-interpolated $\tau$-value between $L$=200\,km \citep[][]{Jewitt19Thermal} and $L$=50,000\,km.  
With the interpolation, we found $\tau <$ 5.96 $\times$ 10$^{-6}$ at $L$=500\,km.   
The difference is only $<$ 0.7\% relative to the measured $\tau$ at $L$=200\,km, which is negligibly small.  
Therefore we used $\tau$\,$<$\,6$\times$10$^{-6}$ for the practical value at $L$=500\,km.}.  
Meanwhile, we also apply $L$=50,000\,km, $\tau$\,$<$\,7$\times$10$^{-9}$ (Table~\ref{optdepth}) 
and the 3$\sigma$ upper limit to $Q_{\rm dust}$\,$\lesssim$\,1\,kg\,s$^{-1}$ when Phaethon was 
at $R_{\rm h}$=1\,au (MJD~58104.2148 in Tables~\ref{W1}) from this {\it WISE} study.  
$U$$\sim$3\,m\,s$^{-1}$ is commonly used for the escape velocity of Phaethon.  
The average separation between particles, $l_a$ (m), is given by $N_1^{-1/3}$.   
The expected number of dust particles encountered at the distance $L$ from Phaethon, 
$N_{\rm enc}$, is estimated by
\begin{equation}  
N_{\rm enc} \simeq N_1\,A_{\rm D^+}\,V_{D^+}\,T_{\rm int},  
\label{enc}
\end{equation}
where $V_{D^+}$ is the flyby speed of DESTINY$^{+}$ and 
$T_{\rm int}$ is its characteristic interaction time. 
We adopt $V_{D^+}$ = 36\,km\,s$^{-1}$ and $T_{\rm int}$ $\sim$ 2$L/V_{D^+}$ to derive $N_{\rm enc}$. 
For the DDA case, $A_{\rm D^+}$ above is replaced with $A_{\rm DDA}$.   
Results are shown in Table~\ref{D+}.  

During the closest approach to Phaethon ($L$$\sim$500\,km), 
DESTINY$^{+}$ is likely to intercept a single millimeter-scale dust particle.  
DDA would encounter two particles of 500$\micron$ in size, though 
these are beyond its sensitivity or scientific scope.   
At far distance from Phaethon ($L$$\sim$50,000\,km), 
a single particle of 200$\micron$ in size may hit DESTINY$^{+}$.  
DDA would also be able to collect six of particles of 
10$\micron$-scale with $N_1$ $\sim$ 10$^{-6}$\,m$^{-3}$.  
The derived DDA detectability is consistent with the results examined by 
the Helios spacecraft measurements and the simulations (IMEX) 
for 13 cometary trails.   
Spatial densities of $\sim$10$\micron$-sized particles 
in the cometary trails of 10$^{-8}$--10$^{-7}$\,m$^{-3}$ 
are detectable with an in-situ instrument \citep[][]{Kruger2020AA}.  

This concept is applicable for the case of the Geminid stream, whereas we have a note.   
Observations of the Geminids find the size range of the 
meteoroids are from nearly millimeter up to 10s of centimeters \citep[e.g.][]{Bla17,Yanagisawa08,Yanagisawa2021}, 
and even smaller particles $<$1\,mm could be present in the stream.  
However, much smaller Geminids with $a \lesssim$10$\micron$ are 
swept away from the stream by the radiation pressure \citep{JewittLi10,Moorhead2021Icar} 
and/or by Poynting-Robertson effect \citep{Ryabova2012MNRASPR}.  
Such smaller Geminids are probably absent in the stream \citep[See also \S3.3 in][]{Jewitt18HST}.
Thus we conclude that the DESTINY$^{+}$ main-unit may be 
intercept a modest number of Geminid meteoroids in the trail.  
DDA is capable of detecting small dust particles, however, 
they are unlikely to be identified as the Geminids.

\subsection{Na Sublimation-driven Mass Production}

The essential puzzle for the Geminid stream 
is its production process from Phaethon.  
Any measured limiting mass-loss rates for Phaethon 
are too small to supply the mass of the Geminid stream (See \S~\ref{W1Q}).  
The rates are almost steady and low over nearly the entire orbit.  
Only at the perihelion the optical brightness is observed to increase suddenly, 
and the mass-loss rate rises sharply (about a factor of 300) for a short time \citep[STEREO, ][]{JewittHsieh2022}.
At perihelion some quasi-episodical dust ejection could have occurred.   
Here, we focus on the perihelion activity of Phaethon 
to find a possible mechanism for producing the Geminid meteoroid stream.  

The optically-observed recurrent perihelion activity is presumed to 
be caused by thermal breakdown (fracture and/or desiccation) of rocks, 
and the release of micron-sized particles at $\sim$3\,kg\,s$^{-1}$, 
the slow ejection speeds of $\sim$3m\,s$^{-1}$, 
and the short tail having mass of $\sim$3$\times$10$^{5}$\,kg are reported 
\citep[][]{JewittLi10,Jewitt12,Jewitt13ApJ,Li_Jewitt2013AJ,HuiLi17}.  
Rotational mass-shedding \citep{Nakano2020ApJ} and 
electrostatic force \citep{Jewitt12,Jewitt_active15,Kimura2022arXiv} are 
also proposed as probable mechanisms. 
But the observed (and modeled) mass-loss rate is tiny and lasted for only $\sim$1\,day 
around perihelion \citep[cf.][]{HuiLi17}.  
Even if it occurred at every perihelion return over the dynamical age $\tau_s$$\sim$1000(s)\,yr, 
the total ejected mass is about 2$\times$10$^{8}$\,kg (with a factor of a few uncertainty from the dynamical age). 
This is more than four to five orders of magnitude too small to be consistent with the stream mass 
$M_s$$\approx$$10^{13}$--$10^{14}$\,kg \citep{HM89,Bla17}, or what would be expected 
if Phaethon were comet \citep[$10^{13}$--$10^{15}$\,kg,][]{Ryabova17}.  
Moreover, the micron-sized particles cannot stay in the stream 
due to their sensitivity to radiation pressure \citep{JewittLi10,Moorhead2021Icar} 
and/or Poynting-Robertson effect \citep{Ryabova2012MNRASPR}.   
Larger particles, such as the near mm$\sim$10s\,cm-sized Geminids in the stream, 
could be simultaneously launched at perihelion but useful data (in longer wavelength) 
and a mechanism for that launching were absent.  

As for the ejection speeds, another consideration is required.   
Dynamical study requires a catastrophic mass production event 
to launch larger particles at high ejection speeds of $\sim$1\,km\,s$^{-1}$, 
proposing 
a comet-like volatile sublimation-driven activity in Phaethon 
at the smallest $q$=0.126\,au about 2,000\,yr ago (during one-time perihelion return) 
\citep{Ryabova2016MNRAS,Ryabova2018MNRASthermal,Ryabova2022PSS}.  
This is modeled by the broad width of the stream at $R_{\rm h}$=1\,au 
from visually observed activities of the Geminid meteor shower \citep{RyabovaRend2018}.  
The high-speed ejection is also suggested by the observations of the 
fine structure of orbits within the Geminid stream (Ji{\v{r}}{\'\i} Borovi{\v{c}}ka, private communication).  
The fatal problem with proposed comet-like activity is that Phaethon is unlikely to contain icy volatiles   
(e.g., H$_2$O, CO, and CO$_2$; see \S~\ref{W2Q}) to make the strong pressure needed to launch the dust.  
Phaethon was unlikely to contain ice 2,000\,yr ago, 
even if it originated from the (icy) inner or outer main-belt asteroids \citep{deLeon10,MacLennan2021Icar}. 
Sublimation-driven ice-loss models find that the NEAs originating from the
inner and outer main-belt would lose ice long before reaching the near-Earth region  \citep{Norbert2020Icar,MacLennan2021Icar}. 
The ice can only possibly be preserved at the polar regions of asteroids having
small axial tilts (obliquity) of $\leq$ 25$^\circ$, or the interiors of asteroids with $D_{\rm e}$ $\ge$ 10\,km \citep{NorbertHenry2018}. 
A situation such as this is improbable for Phaethon given its small size ($D_{\rm e}$$\approx$5--6\,km) 
and high obliquity of $>>$ 25$^\circ$ (a pole orientation of $\lambda_{\rm e}$=$+$85$^\circ$$\pm$13$^\circ$ 
and $\beta_{\rm e}$=$-$20$^\circ$$\pm$10$^\circ$ at ecliptic coordinates \citep{Ansdell2014ApJ}). 
Furthermore, both of the northern and southern hemispheres have experienced intense solar heating at the smallest
$q$=0.126\,au by the pole-shift cycles \citep[][]{Hanus2016AA,MacLennan2021Icar}. 
As such, Phaethon is highly unlikely to contain icy volatiles that could produce the Geminids.  

Recently, \cite{Masiero2021PSJ} proposed that 
sublimation of sodium (Na) may drive the perihelion mass-loss activity of Phaethon. 
Sodium should be or used to be contained in Phaethon 
and the thermal desorption occurs with intense 
solar heating, as observed in the Geminids \citep{KasugaJewitt2019}.  
The near-Sun activity of Phaethon is 
attributable to emissions from neutral Na-D lines (and Fe\,{\sc i} lines), 
as reported by the STEREO coronagraphic observations \citep[][]{Hui2022arXiv}. 
Here, we examine the Na sublimation-driven mass production 
to find its consistency with the formation process of 
the Geminid meteoroid stream.  

We focus on the subsurface of Phaethon at the depth $d$\,$\sim$\,0.05\,m.
There, maximum sodium sublimation pressure ($\lesssim$~1\,Pa) 
and the temperatures $T_d$=580$\sim$770\,K are modeled using 
several different rotational phases at perihelion 
\citep[Figure~(1): Right, ][]{Masiero2021PSJ}\footnote{NIMBUS (Numerical Icy Minor Body evolUtion Simulator) 
is the thermophysical model developed by \cite{Davidsson2022MNRAS}.  
The calculation considers a rotating spherical body consisting of a porous mixture of dust and ice, and tracks 
the internal ice sublimation, vapor condensation, and diffusion of gas and heat in the radial 
and latitudinal directions over time during body rotation \citep[See also,][]{Davidsson2021Icar}.}.
Sublimation of pure sodium is used in the model as a bounding case, 
which is not an unrealistic assumption. 
Sodium is a trace species in rocks and initially would be 
contained in a silicate mineral phase (e.g. feldspar).  
However, the pure phase can be made by segregating from the host 
minerals under severely heated \citep[][]{CapekBorovicka2009} 
or irradiated \citep[][]{Russell1994} conditions like seen in Phaethon 
approaching very near the Sun \citep[See \S~2,][]{Masiero2021PSJ}.

We estimate the maximum size of dust particles to be 
dragged out by sodium sublimation-driven gas.
The critical radius of dust particles, $a_{\rm c}$, 
is related with the saturation partial pressure of 
sodium sublimation over the condensed state, 
$P_{\rm sat}(T_d)$ (in Pa) \citep[Equation~(2),][]{Masiero2021PSJ},  
and given by \citep[][]{Jewitt2002AJ}, 
\begin{equation}
a_c \sim \frac{9 C_D P_{\rm sat} (T_d)}{8 \pi G \rho_d^2 D_e}, 
\end{equation}
where 
$C_{\rm D} \sim$1 is the dimensionless drag coefficient, 
$G$ = 6.67 $\times$ 10$^{-11}$~m$^3$\,kg$^{-1}$\,s$^{-2}$ is the gravitational constant, 
and the same values of $\rho_d$ = 2000\,kg\,m$^{-3}$ and $D_{\rm e}$ = 4.6\,km (from Phaethon) are adopted. 
With $T_d$=580$\sim$770\,K, we obtain $P_{\rm sat}(T_d)$ = 0.003$\sim$0.6\,Pa.
The corresponding critical radius is $a_c$ $\sim$ 0.09--17\,cm, 
comparable to the optically measured size of the Geminid meteoroids 
in lunar impacts \citep[e.g. $\lesssim$\,20\,cm, ][]{Yanagisawa08,Yanagisawa2021}.
By comparison we set $\rho_d$ = 3000\,kg\,m$^{-3}$ estimated from the observations of the Geminid meteors \citep{Baba2009AA} 
to find $a_c$ $\sim$ 0.04--7\,cm, rather consistent to the measured Geminids sizes \cite[cf.][]{Bla17,Yanagisawa08,Yanagisawa2021}.
For the particles around the rotational equator, $a_c$ would be larger 
due to rotation assist by the enhanced angular velocity
\citep[][]{Jewitt2002AJ}, therefore
we consider our estimations as useful lower limit.  
The thermal speed for gas of sodium atoms, $V_{\rm gas}$, is given by 
\citep[Equation (10) of][]{GraykowskiJewitt19}, 
\begin{equation}
V_{\rm gas} (T_d) = \sqrt{\frac{8 k_{\rm B} T_d}{\pi \mu m_{\rm H}}},
\label{vgas} 
\end{equation}
where $k_{\rm B}$ = 1.38$\times$10$^{-23}$\,J\,K$^{-1}$ is the Boltzmann constant, 
$\mu$=22.99 is the molecular weight of sodium, 
$m_{\rm H}$=1.67$\times$10$^{-27}$\,kg is the mass of the hydrogen atom, 
and     
$T_d$ (in K) is the temperature at the depth of 0.05\,m below the Phaethon surface. 
Substituting $T_d$=580$\sim$770\,K, we find $V_{\rm gas}(T_d)$ = 730--840\,m\,s$^{-1}$.  
This is orders of magnitude faster than the other proposed production processes (e.g. thermal breakdown),  
and comparable to the required speeds of $\sim$1\,km\,s$^{-1}$ described by \cite{Ryabova2016MNRAS}.

Next, we estimate the total mass of dust particles ejected by the sodium sublimation-driven activity 
to compare with the Geminid stream mass. 
The total specific sublimation mass-loss rate of sodium at depth $d \sim$0.05\,m, 
($dm/dt$)$_{d}$ in ${\rm kg\,m^{-2}\,s^{-1}}$, 
is given by integrating over temperatures 580 $\lesssim$ $T_d$ $\lesssim$ 770\,K.  
The dust-to-gas (sodium) mass ratio $\sim$1 is applied \citep{Oppenheimer1980}.   
Then we obtain  
\begin{equation}
\left( \frac{dm}{dt} \right)_d = \int~\frac{P_{\rm sat} (T_d) }{V_{\rm gas}(T_d)}~dT_d. 
\label{int_dmdt}
\end{equation}
Adopting $P_{\rm sat}(T_d)$ \citep[][]{Masiero2021PSJ} and Equation~(\ref{vgas}),  
we find $(dm/dt)_{d}$ $\sim$ 0.03 ${\rm kg\,m^{-2}\,s^{-1}}$.  
The mass-loss rate from the entire subsurface 
of Phaethon ($\sim\pi$$D_{\rm e}^2$) corresponds to $\sim$2.0$\times$10$^{6}$\,kg\,s$^{-1}$.  
If a Na-driven perihelion activity likewise lasted for 1\,day each orbital return, 
the total mass over the dynamical age $\tau_s$$\sim$1000(s)\,yr is 
1.2$\times$10$^{14}$\,kg (with a factor of a few uncertainty from the dynamical age).  
This is consistent with the Geminid stream mass $M_s$$\approx$$10^{13}$--$10^{14}$\,kg \citep{HM89,Bla17}, 
and also is comparable to that of the hypothetical cometary nature 
for Phaethon \citep[$M_s$ $\approx$ $10^{13}$--$10^{15}$\,kg,][]{Ryabova17}.  
We note that the calculated Na-driven perihelion mass-loss 
rate is about five or six orders of magnitude larger than the measured mass-loss rates for Phaethon 
in the optical, near- and mid-infrared wavelengths (\S~\ref{W1Q}). 
For example, the STEREO optically estimated the perihelion mass-loss rates 
0.01\,kg\,s$^{-1}$ (= 10$^3$\,kg / 1\,day)  \citep[][]{Hui2022arXiv}, 
up to $\sim$3\,kg\,s$^{-1}$ \citep[][]{JewittHsieh2022}. 
These measured values are associated with the 1$\sim$10$\micron$-sized particles. 
On the other hand, the calculated Na-driven perihelion mass-loss rate is 
presumably related to the near mm$\sim$10s\,cm-sized particles containing more mass. 
We again propose longer wavelength observations aimed at detecting larger 
particles, especially during the perihelion activity of Phaethon.  

We also note that there should be a mechanism to replenish sodium-depleted subsurface 
with sodium-rich layer every return, but unknown yet \citep{JewittHsieh2022}. 
A single perihelion passage obviously results in the depletion of sodium abundance 
down to the depth $d\lesssim$0.05\,m \citep[Figure~(2): Middle left, ][]{Masiero2021PSJ}.  
How had the sodium been supplied into the subsurface for 
the stream lifetime $\tau_s \sim$1000(s)\,yr?   
As a hypothesis, combinations of the suggested active-triggers are considerable.    
Thermal breakdown will be concentrated in thin surface layer 
having thickness similar to the diurnal thermal skin depth 
$d_s \sim$0.1\,m \citep{JewittLi10}, comparable within a factor of two 
in the sodium-driver model case.
The material would be found in thin layer accessible to thermal wave, which would be
disintegrating and peeled off, resulting in a renewal as fresh surface with sodium.   
Still, there remains unknown parameters (e.g. lifetime of thin surface layer) and  
a combination of activity mechanisms may be needed to advance our understanding.   
Future work will investigate a scenario where Phaethon's nature is most likely an asteroid with some of the mass production  
attributable to a past breakup, while the sodium 
can play a contributory role of comet-like activity at perihelion.








\section{Summary} 

We present {\it WISE}/NEOWISE observations 
of the potentially Geminid-associated asteroids,  
Phaethon, 2005~UD and 1999~YC.  
The multi-epoch thermal infrared data 
are obtained in the wavelength bands 
3.4$\micron$ ($W1$), 4.6$\micron$ ($W2$), 
12$\micron$ ($W3$) and 22$\micron$ ($W4$).   
The asteroids appear point-like in all image data.  
We use the data to set limits to the presence of 
dust attributable to the asteroids.  
The decade-long survey data give the following results.

\begin{enumerate}
\item  No evidence of lasting mass-loss was found in the Phaethon image.
          The maximum dust production rate is $Q_{\rm dust}$~$\lesssim$~2\,kg\,s$^{-1}$, having  
          no strong dependency on heliocentric distance at $R_{\rm h}$=1.0--2.3\,au.   
\item  The measured limiting dust production rates are orders of magnitude too small to supply the mass of 
          the Geminid meteoroid stream in the 1000(s)\,yr dynamical age.  
          If Phaethon is the source of the Geminids, the stream mass is likely to be produced episodically, 
          not in steady-state.  
\item  No dust trail is detected around Phaethon when at $R_{\rm h}$=2.3\,au.
	  The corresponding upper limit to the optical depth is $\tau$~$<$~7$\times$10$^{-9}$.  
\item  No co-moving objects were detected. 
	  The limiting radius of a possible source is $r_{\rm e} <$~650\,m 
	  at 42,000\,km distance to Phaethon.
\item  When DESTINY$^{+}$ passes at 50,000\,km distance from Phaethon,  
	  several of 10$\micron$-scale particles would be captured by the dust analyzer, though they most likely will not be identified as the Geminids.    
	  During the flyby phase (at 500\,km distance), two particles of 500$\micron$ in size 
	  are encountered in the instrument.  
\item The 2005~UD image found no extended emission for on-going mass-loss. 
	 The maximum limit to the dust production rate is $Q_{\rm dust}$~$\lesssim$~0.1\,kg\,s$^{-1}$ 
	 at $R_{\rm h}$=1.0--1.4\,au.  
	If dust production were sustained over the $\sim$10,000\,yr dynamical age of 
	the Daytime Sextantid meteoroid stream, the mass of the stream is $\sim$10$^{10}$\,kg.  
	If it were produced by the catastrophic event, the stream mass would be more.     
\item  The 1999~YC data showed no coma at $R_{\rm h}$=1.0--1.6\,au.
	  The maximum mass-loss rate is $Q_{\rm dust}$~$\lesssim$~0.1\,kg\,s$^{-1}$.    
\item   No dust trail associated with 1999~YC was found at $R_{\rm h}$=1.6\,au, corresponding to 
	   the upper limit to the optical depth $\tau$~$<$~2$\times$10$^{-8}$.  
\item   No associated-fragments with radii $r_{\rm e} <$~170\,m were discovered at 24,000\,km distance from 1999~YC.  
\item   The sodium sublimation-driven perihelion activity of Phaethon is expected to eject    
           the near mm$\sim$10s\,cm-sized dust particles at speeds of 730--840\,m\,s$^{-1}$ through gas drag.  
           A mass-loss rate $\sim$2.0$\times$10$^{6}$\,kg\,s$^{-1}$, if allowed to last for 1\,day every return,  
           would deliver about 1.2$\times$10$^{14}$\,kg in 1000(s)\,yr.  
           These processes are compatible with the structure of the Geminid meteoroid stream.   
           For this mechanism to be plausible, though, a sustainable mechanism to resupply sodium-depleted subsurface with sodium-rich minerals is required. 
\end{enumerate}


\begin{acknowledgments}
{\small 
We are grateful to David Jewitt for productive discussion.  
TK thanks 
Mikiya Sato, Chie Tsuchiya, Sunao Hasegawa, Naoya Ozaki, 
Galina Ryabova, Yung Kipreos, Peter Brown, Ji{\v{r}}{\'\i} Borovi{\v{c}}ka, Bj{\"o}rn  J.~R. Davidsson, 
Hideyo Kawakita, and Jun-ichi Watanabe for support.   
This publication makes use of data products from the {\it Wide-field Infrared Survey Explorer}, 
which is a joint project of the University of California, Los Angeles, and 
the Jet Propulsion Laboratory/California Institute of Technology, funded by 
the National Aeronautics and Space Administration. 
Also, this publication makes use of data products from the Near-Earth Object Wide-field 
Infrared Survey Explorer (NEOWISE), which is a joint project of the Jet Propulsion 
Laboratory/California Institute of Technology and the University of Arizona. NEOWISE 
is funded by the National Aeronautics and Space Administration.
This publication uses data obtained from the NASA Planetary Data System (PDS). 
This research has made use of data and services provided by the International Astronomical 
Union's Minor Planet Center.  
This research has made use of the NASA/IPAC Infrared Science Archive, which is 
funded by the National Aeronautics and Space Administration and operated by 
the California Institute of Technology, 
under contract with the National Aeronautics and Space Administration.
We would like to address special thanks to anonymous reviewer and to Maria Womack for scientific editor. 
%
Finally, we express deep gratitudes to Althea Moorhead, 
Margaret Campbell-Brown and 
the LOC members for Meteoroids~2022 (held in Virtual) 
providing opportunity to enhance this study.
}
\end{acknowledgments}

%

\vspace{5mm}
\facilities{WISE, NEOWISE}

\bibliography{ref}
\bibliographystyle{aasjournal}





\clearpage
\input{obslog}


\input{PGCorbits}

\clearpage

\input{Size_pv}
\clearpage

\input{thermalflux}

\clearpage

\input{Teff}

\clearpage

\input{W1}

\clearpage

\input{optdepth}
\clearpage

\input{D+}
\clearpage


%
\clearpage

\clearpage

\clearpage

\clearpage

\clearpage

\clearpage

\clearpage


\end{document}

%% file: obslog.tex
\begin{deluxetable*}{lccrrccccccc}
\tablecaption{{\it WISE}/NEOWISE Observation Log \label{obslog}}
\tablewidth{0pt}
\tabletypesize{\scriptsize}
\tablehead{
\colhead{Object}  &\colhead{\rm MJD\tablenotemark{a}}  & \colhead{UT Date}             & \colhead{RA\tablenotemark{b}}     & \colhead{DEC\tablenotemark{c}}        & \colhead{$R_{\rm h}$\tablenotemark{d}} & \colhead{$\Delta$\tablenotemark{e}} & \colhead{$\alpha$\tablenotemark{f}} & \colhead{$W1$\tablenotemark{g}} & \colhead{$W2$\tablenotemark{h}}& \colhead{$W3$\tablenotemark{i}} & \colhead{$W4$\tablenotemark{j}}\\
                  &                                &                                   &    (deg)                          &      (deg)                        &    (AU)                               &          (AU)                       &         (deg)                       &    (mag)                        &               (mag)            &         (mag)                   &  (mag)                        }
\startdata                                                                            
Phaethon          & 55203.2564$\bullet$            &   2010-Jan-07                     &   9.79                            &     +22.58                        & 2.3171	                            &           2.0755	                  &        25.1                            & 16.995         	          &              15.599	           & 8.736$\pm$0.052	             & 6.155$\pm$0.113 \\
                  & 55203.3888$\bullet$            &   2010-Jan-07                     &   9.81                            &     +22.57                        & 2.3175	                            &           2.0778	                  &        25.1	                           & 16.859$\pm$0.464	          &              15.108	           & 8.700$\pm$0.048	             & 6.050$\pm$0.115 \\
                  & 57035.2144                     &   2015-Jan-13                     &   21.68                           &     +16.90                        & 1.3292	                            &           0.8307	                  &        47.6                         & 14.259$\pm$0.075	          & 11.229$\pm$0.028               &    ---                          &   ---           \\
                  & 57035.2801                     &   2015-Jan-13                     &   21.66                           &     +16.87                        & 1.3283	                            &           0.8309	                  &        47.6                         & 14.467$\pm$0.074	          & 11.361$\pm$0.024               &    ---                          &   ---           \\
                  & 57035.4115                     &   2015-Jan-13                     &   21.62                           &     +16.82                        & 1.3266	                            &           0.8315	                  &        47.7                         & 14.266$\pm$0.076	          & 11.382$\pm$0.027               &    ---                          &   ---           \\
                  & 57035.5429                     &   2015-Jan-13                     &   21.58                           &     +16.76                        & 1.3249	                            &           0.8320	                  &        47.8                         & 14.375$\pm$0.083	          & 11.349$\pm$0.026               &    ---                          &   ---           \\
                  & 57035.6743                     &   2015-Jan-13                     &   21.54                           &     +16.71                        & 1.3232	                            &           0.8325	                  &        47.9                         & 14.300$\pm$0.061	          & 11.288$\pm$0.024               &    ---                          &   ---           \\
                  & 57663.4184                     &   2016-Oct-02                     &   236.56                          &     +84.32                        & 1.0793	                            &           0.4055	                  &        68.0                         & 12.403$\pm$0.025     	          &  9.214$\pm$0.014               &    ---                          &   ---           \\
                  & 57663.5494                     &   2016-Oct-02                     &   239.60                          &     +84.37                        & 1.0813	                            &           0.4057	                  &        67.7                         & 12.478$\pm$0.031     	          &  9.238$\pm$0.014               &    ---                          &   ---           \\
                  & 57663.6149                     &   2016-Oct-02                     &   241.14                          &     +84.40                        & 1.0824	                            &           0.4059	                  &        67.6                         & 12.425$\pm$0.026     	          &  9.233$\pm$0.014               &    ---                          &   ---           \\
                  & 58024.3765                     &   2017-Sep-28                     &   96.73                           &     +33.36                        & 1.9047	                            &           1.6289	                  &        31.7                         & 15.775$\pm$0.197     	          & 14.171$\pm$0.184               &    ---                          &   ---           \\
                  & 58104.2148$\star$              &   2017-Dec-17$\star$              &   3.77                            &     +25.01                        & 1.0068	                            &           0.0691	                  &        68.9                         &  7.782$\pm$0.041$\dagger$       & 4.224$\pm$0.221$\dagger$       &    ---                          &   ---           \\
\hline                                                                                
2005~UD           & 57750.3946                     &   2016-Dec-28                     &   357.82                          &     +47.19                        & 1.3650	                            &           0.7818	                  &        45.1                         & 15.236$\pm$0.118	          & 13.693$\pm$0.104               &    ---                          &   ---           \\
                  & 58383.2977                     &   2018-Sep-22                     &   90.04                           &     +5.90                         & 1.0291	                            &           0.2456	                  &        77.2                         & 13.907$\pm$0.053	          & 10.716$\pm$0.020               &    ---                          &   ---           \\
                  & 58383.4286                     &   2018-Sep-22                     &   89.60                           &     +5.94                         & 1.0312	                            &           0.2447	                  &        76.7                         & 14.199$\pm$0.058	          & 10.917$\pm$0.024               &    ---                          &   ---           \\
                  & 58383.4941                     &   2018-Sep-22                     &   89.38                           &     +5.96                         & 1.0323	                            &           0.2443	                  &        76.5                         & 13.829$\pm$0.058	          & 10.697$\pm$0.021               &    ---                          &   ---           \\
                  & 58821.4792                     &   2019-Dec-04                     &   234.89                          &     +67.36                        & 1.0970	                            &           0.4768	                  &        64.0                         & 15.505$\pm$0.136	          & 12.519$\pm$0.057               &    ---                          &   ---           \\
                  & 58821.6099                     &   2019-Dec-04                     &   235.42                          &     +67.16                        & 1.0950	                            &           0.4759	                  &        64.2                         & 15.255$\pm$0.156 	          & 12.435$\pm$0.051               &    ---                          &   ---           \\
                  & 58821.7408                     &   2019-Dec-04                     &   235.93                          &     +66.96                        & 1.0930	                            &           0.4750	                  &        64.4                         & 15.899$\pm$0.201                & 12.477$\pm$0.055               &    ---                          &   ---           \\
                  & 58821.8716                     &   2019-Dec-04                     &   236.44                          &     +66.76                        & 1.0910	                            &           0.4741	                  &        64.6                         & 15.689$\pm$0.150                & 12.740$\pm$0.054               &    ---                          &   ---           \\
\hline                                                                                
1999~YC           & 55213.2458                     &   2010-Jan-17                     &   27.957                          &     +15.923                       & 1.6242	                            &           1.2137	                  &        37.1                         &            16.303	          & 14.893$\pm$0.317	           & 8.733$\pm$0.055	             & 6.353$\pm$0.155 \\
                  & 55213.5766                     &   2010-Jan-17                     &   27.886                          &     +15.953                       & 1.6208	                            &           1.2157	                  &        37.2                         &            17.094	          &          14.890	           & 9.517$\pm$0.105	             & 7.178$\pm$0.298 \\
                  & 55213.6428$\ast$               &   2010-Jan-17                     &   27.872                          &     +15.959                       & 1.6201	                            &           1.2161	                  &        37.3                         &            16.776	          & 14.760$\pm$0.261	           & 8.953$\pm$0.065	             & 6.650$\pm$0.186 \\
                  & 55213.7089$\ast$               &   2010-Jan-17                     &   27.858                          &     +15.965                       & 1.6194	                            &           1.2165	                  &        37.3                         &         $\diamond$	          &    $\diamond$                  & 8.834$\pm$0.060	             & 6.636$\pm$0.184 \\
                  & 55213.7751$\ast$               &   2010-Jan-17                     &   27.844                          &     +15.971                       & 1.6187	                            &           1.2169	                  &        37.3                         &            16.999	          &          14.841	           & 9.428$\pm$0.099	             & 7.125$\pm$0.287 \\
                  & 55213.8413$\ast$               &   2010-Jan-17                     &   27.830                          &     +15.977                       & 1.6180	                            &           1.2173	                  &        37.3                         &            16.964	          & 14.539$\pm$0.227	           & 9.009$\pm$0.070	             & 6.716$\pm$0.204 \\
                  & 55213.9736                     &   2010-Jan-17                     &   27.802                          &     +15.989                       & 1.6166	                            &           1.2181	                  &        37.4                         &            16.507	          &          14.756	           & 9.174$\pm$0.080	             & 6.853$\pm$0.229 \\
                  & 57738.0234                     &   2016-Dec-16                     &  277.30                           &     +66.69                        & 1.0178	                            &           0.2500	                  &        75.2                         & 13.623$\pm$0.043                & 10.478$\pm$0.02                &    ---                          &   ---           \\
\enddata
\tablecomments{Cases where no measurement was available from NEOWISE a indicated as ``---''.  
Entries without uncertainties indicate a 2$\sigma$ upper limit extracted using the {\it WISE} photometric pipeline.}
\tablenotetext{a}{Modified Julian Date of the mid-point of the observation.}
\tablenotetext{b}{Right ascension (J2000).} 
\tablenotetext{c}{Declination (J2000).} 
\tablenotetext{d}{Heliocentric distance.} 
\tablenotetext{e}{{\it WISE}-centric distance.} 
\tablenotetext{f}{Phase angle.}
\tablenotetext{g}{Magnitude at $W1$.}
\tablenotetext{h}{Magnitude at $W2$.}
\tablenotetext{i}{Magnitude at $W3$.}
\tablenotetext{j}{Magnitude at $W4$.}
\tablenotemark{$\bullet$}{The $W3$ images are combined into Figure~\ref{PhaethonW3}.}
\tablenotemark{$\ast$}{The $W3$ images are combined into Figure~\ref{1999YCW3}.}
\tablenotemark{$\star$}{Phaethon's closest approach date to the Earth.}
\tablenotemark{$\dagger$}{Saturated photometric biases were corrected from the source data of $W1$=7.745$\pm$0.014\,mag and $W2$=2.894$\pm$0.003\,mag following Equation~(\ref{saturated}).}
\tablenotemark{$\diamond$}{Contaminated by a star.}
\end{deluxetable*}

%% file: PGCorbits.tex
\begin{deluxetable*}{lcccccccccc}
\tablecaption{Orbital Properties \label{PGCorbits}}
\tablewidth{0pt}
\tablehead{
\colhead{Object}  &\colhead{$a$\tablenotemark{a}} & \colhead{$e$\tablenotemark{b}} & \colhead{$i$\tablenotemark{c}} & \colhead{$q$\tablenotemark{d}} & \colhead{$\omega$\tablenotemark{e}} & \colhead{$\Omega$\tablenotemark{f}} & \colhead{$Q$\tablenotemark{g}} & \colhead{$P_{\rm orb}$\tablenotemark{h}} & \colhead{$T_{\rm J}$\tablenotemark{i}}\\
                  &   (AU)                        &                                &      (deg)                     &    (AU)               &          (deg)            &         (deg)             &    (AU)    &    (yr)      }
\startdata
Phaethon         & 1.271                & 0.890                & 22.255               & 0.140                 & 322.175                   & 265.223    &   2.403        & 1.43 & 4.509 \\
2005~UD          & 1.275                & 0.872                & 28.664               & 0.163                 & 207.597                   & 19.716     &   2.387        & 1.44 & 4.504 \\	
1999~YC          & 1.422                & 0.830                & 38.217               & 0.241                 & 156.393                   & 64.793     &   2.603        & 1.70 & 4.115 \\
\enddata
\tablecomments{Orbital data were obtained from NASA JPL Small-Body Database (Epoch 2459600.5: 2022-Jan-21)}
\tablenotetext{a}{Semimajor axis.}
\tablenotetext{b}{Eccentricity.} 
\tablenotetext{c}{Inclination.} 
\tablenotetext{d}{Perihelion distance.} 
\tablenotetext{e}{Argument of perihelion.} 
\tablenotetext{f}{Longitude of ascending node.}
\tablenotetext{g}{Aphelion distance.}
\tablenotetext{h}{Orbital period.}
\tablenotetext{i}{Tisserand parameter with respect to Jupiter.  $T_J > 3.08$ for asteroids and $T_J < $ 3.08 for comets, given $a < a_J$ = 5.2\,au \citep{Jewitt_active15}.
  See also, other suggested comet-asteroid thresholds of $T_J = 3.05$ \citep{Tancredi2014Icar} and $T_J = 3.10$ \citep{HsiehHaghighipour2016Icar} \citep[Reviewed in][]{JewittHsieh2022}.}
\end{deluxetable*}

%% file: Size_pv.tex
\begin{deluxetable}{cccc}
\tabletypesize{\scriptsize}
\tablecaption{Effective diameter and geometric visible albedo determined from {\it WISE}/NEOWISE. \label{Size_pv}}
\tablewidth{0pt}
\tablehead{
Object   & \colhead{$D_{\rm e}$\tablenotemark{a}} & \colhead{$p_{\rm v}$\tablenotemark{b}} & Source\\
         &   (km)                                &                                       &       }
\startdata
Phaethon & 4.6$^{+0.2}_{-0.3}$$\ast$              & 0.16$\pm$0.02                         & (1) \\
2005~UD  & 1.2$\pm$0.4                           & 0.14$\pm$0.09                         & (1) \\
1999~YC  & 1.7$\pm$0.2                           & 0.09$\pm$0.03                         & (2) \\
\enddata
\tablenotetext{a}{Effective spherical diameter.}
\tablenotetext{b}{Geometric visible albedo.}
\tablenotemark{$\ast$}{The uncertainty used in this work is $\pm$0.3\,km (see \S~\ref{wise}).}
\tablerefs{(1) \cite{Masiero2019AJ}; (2) \cite{Mainzer2019PDSS}}
\end{deluxetable}


%% file: thermalflux.tex
\begin{deluxetable*}{lccccc}
\tablecaption{Thermal Flux Density \label{thermalflux}}
\tablewidth{0pt}
\tablehead{
\colhead{Object}  &\colhead{$\rm MJD$\tablenotemark{a}}  & \colhead{$W1$\tablenotemark{b}} & \colhead{$W2$\tablenotemark{c}}& \colhead{$W3$\tablenotemark{d}} & \colhead{$W4$\tablenotemark{e}}\\
                  &                                 &      (Jy)                        &               (Jy)             &          (Jy)                   &    (Jy)                        }
\startdata
Phaethon          & 55203.2564                      &    1.87$\times$10$^{-6}$	       &        4.73$\times$10$^{-5}$	& 9.09$\pm$1.73$\times$10$^{-3}$   & 2.93$\pm$0.62$\times$10$^{-2}$ \\
                  & 55203.3888                      & 4.52$\pm$2.10$\times$10$^{-6}$   &        8.44$\times$10$^{-5}$	& 9.40$\pm$1.78$\times$10$^{-3}$   & 3.23$\pm$0.69$\times$10$^{-2}$ \\
                  & 57035.2144                      & 8.33$\pm$1.64$\times$10$^{-5}$   & 4.27$\pm$0.79$\times$10$^{-3}$  &    ---                          &   ---                          \\
                  & 57035.2801                      & 2.15$\pm$0.42$\times$10$^{-5}$   & 3.76$\pm$0.70$\times$10$^{-3}$  &    ---                          &   ---                          \\
                  & 57035.4115                      & 8.35$\pm$1.65$\times$10$^{-5}$   & 3.68$\pm$0.68$\times$10$^{-3}$  &    ---                          &   ---                          \\
                  & 57035.5429                      & 5.01$\pm$1.00$\times$10$^{-5}$   & 3.81$\pm$0.71$\times$10$^{-3}$  &    ---                          &   ---                          \\
                  & 57035.6743                      & 7.30$\pm$1.41$\times$10$^{-5}$   & 4.04$\pm$0.75$\times$10$^{-3}$  &    ---                          &   ---                          \\
                  & 57663.4184                      & 1.11$\pm$0.21$\times$10$^{-3}$   & 2.98$\pm$0.55$\times$10$^{-2}$  &    ---                          &   ---                          \\
                  & 57663.5494                      & 9.49$\pm$1.77$\times$10$^{-4}$   & 2.90$\pm$0.54$\times$10$^{-2}$  &    ---                          &   ---                          \\ 
                  & 57663.6149                      & 1.05$\pm$0.20$\times$10$^{-3}$   & 2.92$\pm$0.54$\times$10$^{-2}$  &    ---                          &   ---                          \\
                  & 58024.3765                      & 3.04$\pm$0.79$\times$10$^{-5}$   & 2.35$\pm$0.59$\times$10$^{-4}$  &    ---                          &   ---                          \\
                  & 58104.2148$\star$               & 0.12$\pm$0.02                   & 3.07$\pm$0.84                   &    ---                          &   ---                          \\
\hline                                             
2005~UD           & 57750.3946                      & 1.24$\pm$1.16$\times$10$^{-4}$   & 4.47$\pm$4.18$\times$10$^{-4}$  &    ---                          &   ---                          \\
                  & 58383.2977                      & 4.21$\pm$3.92$\times$10$^{-4}$   & 7.69$\pm$7.16$\times$10$^{-3}$  &    ---                          &   ---                          \\
                  & 58383.4286                      & 2.80$\pm$2.61$\times$10$^{-4}$   & 6.36$\pm$5.92$\times$10$^{-3}$  &    ---                          &   ---                          \\
                  & 58383.4941                      & 4.60$\pm$4.29$\times$10$^{-4}$   & 7.82$\pm$7.28$\times$10$^{-3}$  &    ---                          &   ---                          \\
                  & 58821.4792                      & 7.45$\pm$6.99$\times$10$^{-5}$   & 1.41$\pm$1.31$\times$10$^{-3}$  &    ---                          &   ---                          \\
                  & 58821.6099                      & 1.08$\pm$1.02$\times$10$^{-4}$   & 1.53$\pm$1.43$\times$10$^{-3}$  &    ---                          &   ---                          \\
                  & 58821.7408                      & 3.62$\pm$3.43$\times$10$^{-5}$   & 1.47$\pm$1.37$\times$10$^{-3}$  &    ---                          &   ---                          \\
                  & 58821.8716                      & 5.46$\pm$5.14$\times$10$^{-5}$   & 1.15$\pm$1.07$\times$10$^{-3}$  &    ---                          &   ---                          \\
\hline                                             
1999~YC           & 55213.2458                      &       4.11$\times$10$^{-5}$      & 1.40$\pm$0.70$\times$10$^{-4}$	& 1.04$\pm$0.42$\times$10$^{-2}$  & 2.44$\pm$1.05$\times$10$^{-2}$ \\
                  & 55213.5766                      &       1.58$\times$10$^{-5}$      &      1.40$\times$10$^{-4}$	& 5.07$\pm$2.11$\times$10$^{-3}$  & 1.14$\pm$0.56$\times$10$^{-2}$ \\
                  & 55213.6428                      &       2.38$\times$10$^{-5}$      & 1.59$\pm$0.75$\times$10$^{-4}$	& 8.53$\pm$3.49$\times$10$^{-3}$  & 1.86$\pm$0.82$\times$10$^{-2}$ \\
                  & 55213.7089                      &          $\diamond$             &        $\diamond$       	& 9.51$\pm$3.89$\times$10$^{-3}$  & 1.88$\pm$0.83$\times$10$^{-2}$ \\
                  & 55213.7751                      &       1.81$\times$10$^{-5}$      &      1.47$\times$10$^{-4}$	& 5.51$\pm$2.29$\times$10$^{-3}$  & 1.20$\pm$0.58$\times$10$^{-2}$ \\
                  & 55213.8413                      &       1.89$\times$10$^{-5}$      & 1.96$\pm$0.89$\times$10$^{-4}$	& 8.11$\pm$3.33$\times$10$^{-3}$  & 1.75$\pm$0.78$\times$10$^{-2}$ \\
                  & 55213.9736                      &       3.28$\times$10$^{-5}$      & 1.60$\times$10$^{-4}$	        & 6.97$\pm$2.87$\times$10$^{-3}$  & 1.54$\pm$0.70$\times$10$^{-2}$ \\
                  & 57738.0234                      &  7.38$\pm$3.01$\times$10$^{-4}$  & 9.60$\pm$3.90$\times$10$^{-3}$	        &    ---                          &   ---                          \\
\enddata
\tablecomments{Reflected sunlight has been removed from the measured magnitudes in Table~\ref{obslog}.   
Cases where no measurement is available in NEOWISE are marked as ``---''.  
Entries without uncertainties are 2$\sigma$ upper limits.  
}
\tablenotetext{a}{Modified Julian Date of the mid-point of the observation.}
\tablenotetext{b}{Thermal flux density at $W1$.}
\tablenotetext{c}{Thermal flux density at $W2$.}
\tablenotetext{d}{Thermal flux density at $W3$.}
\tablenotetext{e}{Thermal flux density at $W4$.}
\tablenotemark{$\star$}{Phaethon's closest approach date to the Earth (2017-Dec-17).}
\tablenotemark{$\diamond$}{Contaminated by a star.}
\end{deluxetable*}

%% file: Teff.tex
\begin{deluxetable*}{lcccccccccc}
\tablecaption{Temperature \label{Teff}}
\tablewidth{0pt}
\tablehead{
\colhead{Object}  &\colhead{MJD\tablenotemark{a}}  & \colhead{$W3/W4$\tablenotemark{b}} & \colhead{$T_{\rm bb}$\tablenotemark{c}}     & \colhead{$T_{\rm eff}$\tablenotemark{d}} &   \colhead{$T_{\rm ss}$\tablenotemark{e}}          \\
                  &                                &                                    &    (K)                                     &    (K)                                  &     (K)                                            }
\startdata                                                                                                                                                                                                                   
Phaethon          & 55203.2564                      &       0.31$\pm$0.09                &   183                                      &   184$^{+18}_{-19}$                      &    219$^{+21}_{-23}$                               \\
                  & 55203.3888                      &       0.29$\pm$0.08                &   183                                      &   180$^{+16}_{-18}$                      &    214$^{+19}_{-21}$                               \\
                  & 57035.2144                      &     ---$^\dagger$                   &   241                                      &   236$^\ddagger$                         &    281                                  \\
                  & 57035.2801                      &     ---$^\dagger$                   &   241                                      &   236$^\ddagger$                         &    281                                  \\
                  & 57035.4115                      &     ---$^\dagger$                   &   241                                      &   236$^\ddagger$                         &    281                                  \\
                  & 57035.5429                      &     ---$^\dagger$                   &   242                                      &   237$^\ddagger$                         &    282                                  \\
                  & 57035.6743                      &     ---$^\dagger$                   &   242                                      &   237$^\ddagger$                         &    282                                  \\
                  & 57663.4184                      &     ---$^\dagger$                   &   268                                      &   263$^\ddagger$                         &    312                                  \\
                  & 57663.5494                      &     ---$^\dagger$                   &   267                                      &   262$^\ddagger$                         &    311                                  \\
                  & 57663.6149                      &     ---$^\dagger$                   &   267                                      &   262$^\ddagger$                         &    311                                  \\
                  & 58024.3765                      &     ---$^\dagger$                   &   201                                      &   197$^\ddagger$                         &    234                                  \\
                  & 58104.2148$\star$               &     ---$^\dagger$                   &   277                                      &   271$^\ddagger$                         &    323                                  \\
\hline                                                                                                                                                                                                           
2005~UD           & 57750.3946                      &     ---$^\dagger$                   &   238                                      &   233$^\ddagger$                         &    277                                  \\
                  & 58383.2977                      &     ---$^\dagger$                   &   274                                      &   269$^\ddagger$                         &    319                                  \\
                  & 58383.4286                      &     ---$^\dagger$                   &   274                                      &   269$^\ddagger$                         &    319                                  \\
                  & 58383.4941                      &     ---$^\dagger$                   &   274                                      &   269$^\ddagger$                         &    319                                  \\
                  & 58821.4792                      &     ---$^\dagger$                   &   265                                      &   260$^\ddagger$                         &    309                                  \\
                  & 58821.6099                      &     ---$^\dagger$                   &   266                                      &   261$^\ddagger$                         &    310                                  \\
                  & 58821.7408                      &     ---$^\dagger$                   &   266                                      &   261$^\ddagger$                         &    310                                  \\
                  & 58821.8716                      &     ---$^\dagger$                   &   266                                      &   261$^\ddagger$                         &    310                                  \\
\hline                                                                                                                                                                                                                                       
1999~YC           & 55213.2458                      &     0.43$\pm$0.25                  &   218                                      &   208$^{+49}_{-55}$                      &    247$^{+58}_{-65}$                              \\
                  & 55213.5766                      &     0.44$\pm$0.28                  &   218                                      &   209$^{+54}_{-59}$                      &    249$^{+64}_{-70}$                              \\
                  & 55213.6428                      &     0.46$\pm$0.28                  &   218                                      &   214$^{+53}_{-59}$                      &    254$^{+63}_{-70}$                              \\
                  & 55213.7089                      &     0.51$\pm$0.31                  &   218                                      &   223$^{+60}_{-64}$                      &    265$^{+71}_{-76}$                              \\
                  & 55213.7751                      &     0.46$\pm$0.29                  &   219                                      &   214$^{+55}_{-61}$                      &    254$^{+65}_{-72}$                              \\ 
                  & 55213.8413                      &     0.46$\pm$0.28                  &   219                                      &   214$^{+53}_{-59}$                      &    254$^{+63}_{-70}$                              \\ 
                  & 55213.9736                      &     0.45$\pm$0.28                  &   219                                      &   212$^{+53}_{-59}$                      &    252$^{+63}_{-70}$                              \\
                  & 57738.0234                      &     ---$^\dagger$                   &   276                                      &   270$^\ddagger$                         &    322                                  \\ 
\enddata
\tablenotetext{a}{Modified Julian Date of the mid-point of the observation.}
\tablenotetext{b}{Flux density ratio from Table~\ref{thermalflux}.} 
\tablenotetext{c}{Blackbody temperature calculated from $T_{\rm bb}$=278$\cdot$$R_{h}^{-0.5}$, where $R_{h}$(au) is from Table~\ref{obslog}.} 
\tablenotetext{d}{Effective temperature.}
\tablenotetext{e}{Subsolar temperature calculated from $T_{\rm ss}$=$\chi^{0.25}$$\cdot$$T_{\rm eff}$, where $\chi$=2 for observable asteroid hemisphere. }
\tablenotemark{$\dagger$}{No data available in NEOWISE.}
\tablenotemark{$\ddagger$}{Assumed value calculated from $T_{\rm eff}$=0.98$\cdot$$T_{\rm bb}$.}
\tablenotemark{$\star$}{Phaethon's closest approach date to the Earth (2017-Dec-17).}
\end{deluxetable*}

%% file: W1.tex
\begin{deluxetable*}{lcccc}
\tablecaption{Dust Production Rate \label{W1}}
\tablewidth{0pt}
\tablehead{
\colhead{Object}  &\colhead{$\rm MJD$\tablenotemark{a}} & \colhead{$M_{\rm d}$\tablenotemark{b}} & \colhead{$Q_{\rm dust}$\tablenotemark{c}}  \\ 
                  &                                 &            (kg)                       &              (kg s$^{-1}$)                 } 
\startdata
Phaethon          & 55203.2564                      &         $<$~6.9$\times$10$^{5}$       &       $<$~0.18                           \\ 
                  & 55203.3888                      &      2.9$\pm$1.5$\times$10$^{6}$      &       0.77$\pm$0.40                       \\ 
                  & 57035.2144                      &           n/a                         &               n/a                        \\ 
                  & 57035.2801                      &           n/a                         &               n/a                        \\ 
                  & 57035.4115                      &           n/a                         &               n/a                        \\ 
                  & 57035.5429                      &           n/a                         &               n/a                        \\ 
                  & 57035.6743                      &           n/a                         &               n/a                        \\ 
                  & 57663.4184                      &           n/a                         &               n/a                        \\ 
                  & 57663.5494                      &           n/a                         &               n/a                        \\ 
                  & 57663.6149                      &           n/a                         &               n/a                        \\ 
                  & 58024.3765                      &       2.0$\pm$0.7$\times$10$^{6}$     &       0.68$\pm$0.24                      \\ 
                  & 58104.2148$\star$               &       2.7$\pm$2.2$\times$10$^{4}$     &       0.22$\pm$0.18                      \\ 
\hline                                             
2005~UD           & 57750.3946                      &    1.0$^{+1.4}_{-1.0}$$\times$10$^{5}$  &    2.4$^{+3.4}_{-2.4}$$\times$10$^{-2}$    \\ 
                  & 58383.2977                      &    1.1$^{+6.8}_{-1.1}$$\times$10$^{3}$  &   0.8$^{+5.1}_{-0.8}$$\times$10$^{-3}$     \\ 
                  & 58383.4286                      &           n/a                         &              n/a                         \\
                  & 58383.4941                      &    1.5$^{+7.1}_{-1.5}$$\times$10$^{3}$  &    1.1$^{+5.2}_{-1.1}$$\times$10$^{-3}$    \\
                  & 58821.4792                      &     1.8$^{+61}_{-1.8}$$\times$10$^{2}$  &    6.9$^{+230}_{-6.9}$$\times$10$^{-5}$   \\
                  & 58821.6099                      &    2.2$^{+8.0}_{-2.2}$$\times$10$^{3}$  &    8.5$^{+31}_{-8.5}$$\times$10$^{-4}$   \\
                  & 58821.7408                      &           n/a                         &               n/a                        \\
                  & 58821.8716                      &           n/a                         &               n/a                        \\ 
\hline                                             
1999~YC           & 55213.2458                      &        $<$~5.8$\times$10$^{5}$        &        $<$~8.8$\times$10$^{-2}$           \\ 
                  & 55213.5766                      &        $<$~1.9$\times$10$^{5}$        &       $<$~2.9$\times$10$^{-2}$	       \\ 
                  & 55213.6428                      &        $<$~2.0$\times$10$^{5}$        &        $<$~3.0$\times$10$^{-2}$           \\ 
                  & 55213.7089                      &              $\diamond$               &           $\diamond$         	       \\ 
                  & 55213.7751                      &        $<$~1.5$\times$10$^{5}$        &          $<$~2.3$\times$10$^{-2}$	       \\ 
                  & 55213.8413                      &        $<$~1.6$\times$10$^{5}$        &         $<$~2.4$\times$10$^{-2}$          \\ 
                  & 55213.9736                      &        $<$~3.2$\times$10$^{5}$        &         $<$~4.8$\times$10$^{-2}$	       \\ 
                  & 57738.0234                      &           n/a                         &               n/a                        \\ 
\enddata
\tablecomments{The 2$\sigma$ upper limit is represented with ``$<$''.  
A result of a negative value is expressed as ``n/a''.}
\tablenotetext{a}{Modified Julian Date of the mid-point of the observation.}
\tablenotetext{b}{Dust mass from Equation~(\ref{mass}).}
\tablenotetext{c}{Dust production rate from Equation~(\ref{Qdust}).}
\tablenotemark{$\star$}{Phaethon's closest approach date to the Earth (2017-Dec-17).}
\tablenotemark{$\diamond$}{Contaminated by a star.}
\end{deluxetable*}

%% file: optdepth.tex
\begin{deluxetable}{cccccccc}
\tabletypesize{\scriptsize}
\tablecaption{Optical Depth \label{optdepth}}
\tablewidth{0pt}
\tablehead{
Object   &    $I_{\rm n}$\tablenotemark{a} & $I_{\rm d}$\tablenotemark{b}  & $C_{\rm n}$\tablenotemark{c}     & $C_{\rm d}$\tablenotemark{d}  &  $\tau$\tablenotemark{e} & $<$\,$\tau$\tablenotemark{f} \\
         &    (Jy)                        & (Jy)                         &  (km$^2$)                       & (km$^2$)                               &          &  }
\startdata
Phaethon & 8.5$\pm$0.6$\times$10$^{-3}$ & 3.7$\pm$1.0$\times$10$^{-6}$    & 16.6$\pm$3.9                    & 7.2$\pm$2.6$\times$10$^{-3}$ & 3.2$\pm$1.2$\times$10$^{-9}$   & 7$\times$10$^{-9}$  \\
1999~YC  & 2.2$\pm$0.1$\times$10$^{-2}$ & 5.4$\pm$2.7$\times$10$^{-5}$    &  2.3$\pm$0.5                    & 5.6$\pm$3.1$\times$10$^{-3}$ & 7.2$\pm$4.0$\times$10$^{-9}$   & 2$\times$10$^{-8}$   \\
\enddata
\tablecomments{From the $W3$-band.}
\tablenotetext{a}{Measured flux density scattered by the asteroid's cross sectional area.}
\tablenotetext{b}{Measured flux density scattered by the dust particles in the trail.}
\tablenotetext{c}{Cross section of asteroid calculated from Table~\ref{Size_pv}.}
\tablenotetext{d}{Cross section of dust particles in the trail (Equation~\ref{ratio}).}
\tablenotetext{e}{Optical depth.}
\tablenotetext{f}{Upper limit to the optical depth ($>$~3$\sigma$).} 
\end{deluxetable}

%% file: D+.tex
\begin{deluxetable}{cccccccc}
\tabletypesize{\scriptsize}
\tablecaption{Expectations for Dust Particle Encounters with DESTINY$^+$ and the Dust Analyzer (DDA) \label{D+} on the Phaethon Mission}
\tablewidth{0pt}
\tablehead{
$L$\tablenotemark{a}&    Unit     & $a$\tablenotemark{b} & $M$\tablenotemark{c}  & $N_1$\tablenotemark{d} & $l_a$\tablenotemark{e}  &  $N_{\rm enc}$\tablenotemark{f}\\
  (km)              &             &    ($\micron$)       &    (kg)               & (m$^{-3}$)            &         (m)                     &                                       }
\startdata                                                                     
500                 & DESTINY$^+$ &      2300            &  1.0$\times$10$^{-4}$  & 7.2$\times$10$^{-7}$   &         110                     &          1                 \\
\smallskip                                                                     
                    &    DDA      &      260             &  1.5$\times$10$^{-7}$  & 5.7$\times$10$^{-5}$    &        26                      &          2                 \\
\hline                                                                         
50,000              & DESTINY$^+$ &      80              &  4.3$\times$10$^{-9}$  & 7.0$\times$10$^{-9}$    &        520                     &          1                 \\
                    &    DDA      &      9               &  6.1$\times$10$^{-12}$ & 1.7$\times$10$^{-6}$    &        84                      &          6                 \\
\enddata
\tablecomments{The characteristic interaction time $T_{\rm int}$ is 28\,seconds at $L$=500\,km and 2800\,seconds at $L$=50,000\,km, respectively.}
\tablenotetext{a}{Distance from Phaethon.}
\tablenotetext{b}{Radius of a dust particle.}
\tablenotetext{c}{Mass of a dust particle.}
\tablenotetext{d}{Number density of dust particles.}
\tablenotetext{e}{Average separation distance between particles ($= N{_1}^{-\frac{1}{3}})$.}
\tablenotetext{f}{Expected number of dust particles to be encountered.}
\end{deluxetable}